\documentclass[3p,,preprint,12pt]{elsarticle}

\usepackage{lineno,hyperref}
\modulolinenumbers[1]
\usepackage{float}
\usepackage{amsmath}
\usepackage{amssymb}
\usepackage{pgfplots}
\usepackage{subcaption}

\usepackage{array}
\newcolumntype{P}[1]{>{\centering\arraybackslash}p{#1}}
 
\usepackage{setspace}
\biboptions{sort&compress}
\AtBeginDocument{\hypersetup{pdfborder={0 0 0}}}
\begin{document}

\begin{frontmatter}

%\title{\textbf{Inversion of composite material properties and layup sequence type using ultrasonic guided waves and CNN with dual-branch feature fusion}}

%\title{\textbf{Composite material characterization using ultrasonic guided waves and CNN with dual-branch feature fusion}}

\title{\textbf{Inverse characterization of composites using guided waves and convolutional neural networks with dual-branch feature fusion}}

%\title{\textbf{Composite material characterization using ultrasonic guided waves and deep convolutional networks}}


\author[a]{Mahindra Rautela \corref{contrib-0}}
\ead{mrautela@iisc.ac.in}\cortext[contrib-0]{Corresponding author}

\author[b]{Armin Huber}
\ead{armin.huber@dlr.de}

\author[c]{J. Senthilnath}
\ead{j\_senthilnath@i2r.a-star.edu.sg}

\author[a]{S. Gopalakrishnan}
\ead{krishnan@iisc.ac.in}

\address[a]{Department of Aerospace Engineering, Indian Institute of Science, Bangalore, India}
\address[b]{Center for Lightweight Production Technology, German Aerospace Center (DLR), Augsburg, Germany}
\address[c]{Institute for Infocomm Research, A*STAR, Singapore}

%---------------------- begin ABSTRACT
\begin{abstract}
In this work, ultrasonic guided waves and a dual-branch version of convolutional neural networks are used to solve two different but related inverse problems, i.e., finding layup sequence type and identifying material properties. In the forward problem, polar group velocity representations are obtained for two fundamental Lamb wave modes using the stiffness matrix method. For the inverse problems, a supervised classification-based network is implemented to classify the polar representations into different layup sequence types (inverse problem - 1) and a regression-based network is utilized to identify the material properties (inverse problem - 2). 
\end{abstract}

%---------------------- begin KEYWORDS
\begin{keyword}
	Material characterization \sep Property Identification \sep Inverse problem \sep Guided waves \sep Deep learning \sep Dual-branch CNN 
\end{keyword}

\end{frontmatter}

%\linenumbers
%----------------------begin INTRODUCTION
\section{Introduction}\label{introduction}
Composite materials have revolutionized different industries due to their high strength-to-weight ratio and improved manufacturing techniques at lower costs. However, due to uncertainties involved at different stages, beginning from the design phase to the end of the material's service life, accurate determination of material properties presents unprecedented challenges. Therefore, characterization of composite materials is necessary not only for non-destructive property measurements but also for real-time material degradation aspects. Different dynamics based non-destructive methods are adopted for characterization of composite materials like ultrasonic bulk waves \cite{kline2017nondestructive,paterson2018elastic,sevenois2018multiscale,nelson2018ply,martens2019characterization,nelson2019fibre}, vibration-based \cite{tam2017identification,tam2018inverse} and ultrasonic guided waves \cite{balasubramaniam1998inversion,hosten2001identification,vishnuvardhan2007genetic,cui2019identification,kudela2020elastic}. Among these, ultrasonic guided waves (UGW) based technique has gained a tremendous amount of attention in recent years for non-destructive evaluation (NDE) and structural health monitoring (SHM) of composite structures \cite{kundu2003ultrasonic,giurgiutiu2007structural,boller2009encyclopedia,gopalakrishnan2011computational}. UGW are sensitive to material properties and offer advantages in terms of traveling larger distances with minimum energy loss. It makes the inspection process rapid and less labor-intensive \cite{mitra2016guided}. Despite such merits, modeling and experimentation with guided wave propagation in composite materials are complex and much more involved. It makes the inverse problem of material characterization complicated, which requires proper attention \cite{balasubramaniam1998inversion}. Many efforts have been dedicated in recent years to solve this problem.

Balasubramaniam \cite{balasubramaniam1998inversion} has used a multi-parameter minimization-based objective function where a genetic algorithm (GA) based optimization is applied to find stiffness constants and ply-layup sequence using fundamental Lamb wave modes. It is reported that inverting the ply-layup sequence is more cumbersome and time-consuming, even with four stacking sequences. Hosten et al. \cite{hosten2001identification} have used experimentally determined phase velocities of Lamb wave modes over a frequency range to calculate material properties. In this methodology, a Newton-Raphson scheme and a simplex algorithm are utilized to minimize the determinant of the Thomson/Haskell matrix. It is concluded that multi-modal phase velocity information is essential for the accurate determination of material properties. Vishnuvardhan et al. \cite{vishnuvardhan2007genetic} have utilized the GA-based approach to predict the elastic properties of three different composites using multi-directional Lamb wave velocities information coming from a single-transmitter-multiple-receiver array. Cui and Scalea \cite{cui2019identification} have used a semi-analytical finite element (SAFE) model to solve the forward problem with three guided wave modes. They have investigated a property inversion scheme based on matching phase velocity dispersion curves of relevant guided modes using simulated annealing (SA) based optimization algorithm. Kudela et al. \cite{kudela2020elastic} have used the SAFE model to determine dispersion curves of Lamb wave modes in the forward problem, whereas GA is used as the inversion scheme. Chen et. al. \cite{chen2021high} have used rotation invariant technique to extract dispersion curve of the Lamb waves and particle swarm optimization (PSO) technique is utilized to estimate the elastic constants for isotropic and transversely isotropic plates. In addition to non-destructive property measurements and real-time property degradation of composites, guided waves along with different inverse algorithms are also used for biomedical applications like bone characterization \cite{bochud2016genetic,minonzio2020automatic,li2020deep}.

Most of the abovementioned research works have used different heuristics-based global optimization algorithms as property inversion schemes. However, these schemes face disadvantages in terms of computational time, large-scale automation, in-situ predictions, adaptability, and generalization. Along with these demerits, these heuristic methods do not guarantee a global optimum. On the other hand, the learning-based technique uses less computational time.  Once the networks are trained, it only requires testing the extracted knowledge, which is reasonably faster than optimizing each set. Also, learning-based models are feasible in real-time deployment.
% On the other hand, improved gradient descent-based local optimization scheme along with the learning ability of neural networks guarantee local optimum and can overcome the drawbacks of global inversion schemes.

The author's recent works \cite{rautela2020ultrasonicIEEE,gopalakrishnan2020deep} have successfully implemented convolutional and recurrent neural networks for the first-time to inversely map two fundamental guided wave modes to the material properties of a transversely-isotropic composite material. In this work, a reduced-order spectral finite element method (SFEM) is used as a forward model. It is observed that the supervised deep learning-based networks are capable of automatic feature extraction from the guided wave modes and learning the inverse relationship between guided wave modes and material properties. Apart from neural network's learning ability, they have also solved the problems with reduced computational time, fast in-situ predictions, and the possibility of large-scale automation with modern graphical processing units (GPU). However, the inverse formulation of property identification is limited by its uniqueness and insufficient forward information to predict the material properties with accuracy. It is highlighted in the literature \cite{balasubramaniam1998inversion,hosten2001identification,vishnuvardhan2007genetic,cui2019identification,kudela2020elastic} that multi-modal, multi-directional ultrasonic guided waves information over a broadband frequency range can enhance the prediction capabilities of inverse models.

In this work, stiffness matrix method (SMM) \cite{rokhlin2002stable,wang2001stable} along with group velocity calculation \cite{giurgiutiu2021stress} is utilized as a forward model. It is used to collect multi-modal polar group-velocity information of ultrasonic guided waves in the form of polar representations over a frequency range. The inverse problem of material characterization is formulated in two different but related problems. In the first inverse problem, a dual-branch version of the convolutional neural network (CNN) based classification model is used to classify the polar representations into three classes of layup sequence types (quasi-isotropic, unidirectional and cross-ply laminates). In the next inverse problem, another dual-branch CNN-based regression model is used to map polar representations to material properties of the transversely-isotropic composite laminate. Both of the problems assume constant geometrical properties. It is seen in the literature that most of the attempts have been put forward to predict the material properties. However, little to no work is available to find the stacking sequence because of the complexity associated with the problem. In Ref.~\cite{balasubramaniam1998inversion}, it is highlighted that inverting the ply-layup sequence is highly non-linear, cumbersome, and time-taking. Therefore, instead of predicting the exact stacking sequence, the complexity of this problem is reduced by classifying the stacking sequences into three important and popular layup classes. This approach may not solve the entire problem of inverting the ply-layup sequence, but the methodology can give some additional and important insights into the overall inverse problem of composite material characterization. 

The paper is presented as follows: Section-\ref{sec:theory} contains theoretical background of the forward (SMM for UGW propagation) and inverse models (Dual-branch CNN). Section-\ref{sec:sensitivity} presents sensitivity analysis of the forward model. Section-\ref{sec:train} contains the training strategy for the networks. Section-\ref{sec:findply} and \ref{sec:findproperties} present application of dual-branch CNN for identification of layup sequence type and material properties, respectively. Testing results and comparisons are discussed in Section-\ref{sec:results} and the paper is concluded in Section-\ref{sec:conclusion}.

% -------------------- Begin NEW SECTION
\section{Theoretical Background}\label{sec:theory}
\subsection{Forward model: Stiffness matrix method for UGW propagation}\label{ssec:forward}
The stiffness matrix method \cite{rokhlin2002stable,wang2001stable} and group velocity calculation routine \cite{giurgiutiu2021stress} is used to solve the forward problem. By their nature, ultrasonic Lamb waves evolve through the propagation and superposition of bulk waves in waveguides. In the case of constructive superposition, a case of resonance and hence a propagating Lamb wave is obtained. Therefore, finding the propagating direction of the bulk waves is of fundamental importance \cite{rauter2018wave}. For guided wave modeling, bulk waves are referred to as partial waves, and the SMM is a so-called partial wave method, which allows finding modal solutions by superimposing the contributing bulk waves in every layer. A maximum of six bulk waves can propagate in every layer of an anisotropic waveguide, namely upward and downward propagating quasi-longitudinal waves (L$^-$, L$^+$), fast quasishear waves (S$^-_\mathrm{fast}$, S$^+_\mathrm{fast}$), and slow quasishear waves (S$^-_\mathrm{slow}$, S$^+_\mathrm{slow}$). Due to the presence of boundaries and interfaces in a waveguide, guided waves are dispersive in nature, i.e., the phase and group velocities are frequency-dependent. In order to obtain dispersion diagrams, the problem is solved for extended frequency ranges. For anisotropic composites, the phase and group velocities are also dependent on the propagation direction within the composite laminates. Therefore, polar dispersion diagrams are calculated for wave propagation angles ranging from $\varPhi=0\,^\circ$ to $360\,^\circ$ with certain increments. 

Acoustic field theory in solids is based on three equations i.e., two field equations (strain-displacement relation and equation of motion) and one elastic constitutive equation. In cartesian crystallographic coordinate system $x'_i=(x'_1,x'_2,x'_3)$, the relation between particle displacements $u'_i$ is related to strain fields $\varepsilon'_{kl}$ via strain-displacement relation as given in Eq.~(\ref{eq.01}).
\begin{equation}\label{eq.01}
	\varepsilon'_{kl}=\frac{1}{2}\left(\frac{\partial u'_{l}}{\partial x'_{k}}+\frac{\partial u'_{k}}{\partial x'_{l}}\right).
\end{equation}

\noindent The equation of motion is given by Eq.~(\ref{eq.02}).
\begin{equation}\label{eq.02}
\frac{\partial\sigma'_{ij}}{\partial x'_j}=\rho\frac{\partial^2 u'_{i}}{\partial t^2},
\end{equation}

where $\sigma'_{ij}$ are the stress field components and $\rho$ is the density of the solid. The elastic constitutive equation, also known as Hooke's law, is stated in Eq.~(\ref{eq.03}).
\begin{equation}\label{eq.03}
\sigma'_{ij}=c'_{ijkl}\varepsilon'_{kl},\hspace{5mm}\varepsilon'_{ij}=s'_{ijkl}\sigma'_{kl}
\end{equation}

It relates the stress field components with the strain field components via the stiffness and compliance tensors $c'_{ijkl}$ and $s'_{ijkl}$. Since $\sigma'_{ij}$ and $\varepsilon'_{kl}$ are symmetric, i.e., $\sigma'_{ij}=\sigma'_{ji}$ and $\varepsilon'_{kl}=\varepsilon'_{lk}$, and because of strain energy considerations, the number of independent elements in $c'_{ijkl}$ reduces from eighty-one to twenty-one (maximum number of independent elastic constants for an anisotropic material). In transversely isotropic materials such as uni-directional fiber-epoxy layers, the number is reduced further to only five. In this situation, it is convenient to write Hooke's law in matrix form using Eq.~(\ref{eq.04}).

%, where 1$\rightarrow$11, 2$\rightarrow$22, 3$\rightarrow$33, 4$\rightarrow$23, 5$\rightarrow$13, and 6$\rightarrow$12. 

\begin{equation}\label{eq.04}
\begin{bmatrix}
	\sigma'_{11}\\
	\sigma'_{22}\\
	\sigma'_{33}\\
	\sigma'_{23}\\
	\sigma'_{13}\\
	\sigma'_{12}\\
\end{bmatrix}=
\begin{bmatrix}
	C'_{11} & C'_{12} & C'_{12} & 0 & 0 & 0\\
	& C'_{22} & C'_{23} & 0 & 0 & 0\\
	& & C'_{22} & 0 & 0 & 0\\
	& & & \frac{C'_{22}-C'_{23}}{2} & 0 & 0\\
	& & \mathrm{sym} & & C'_{55} & 0\\
	& & & & & C'_{55}\\    
\end{bmatrix}
\begin{bmatrix}
	\varepsilon'_{11}\\
	\varepsilon'_{22}\\
	\varepsilon'_{33}\\
	2\varepsilon'_{23}\\
	2\varepsilon'_{13}\\
	2\varepsilon'_{12}\\
\end{bmatrix}.
\end{equation}

\begin{figure}[t]	
\parbox{.49\textwidth}{
	\includegraphics[width=.47\textwidth]{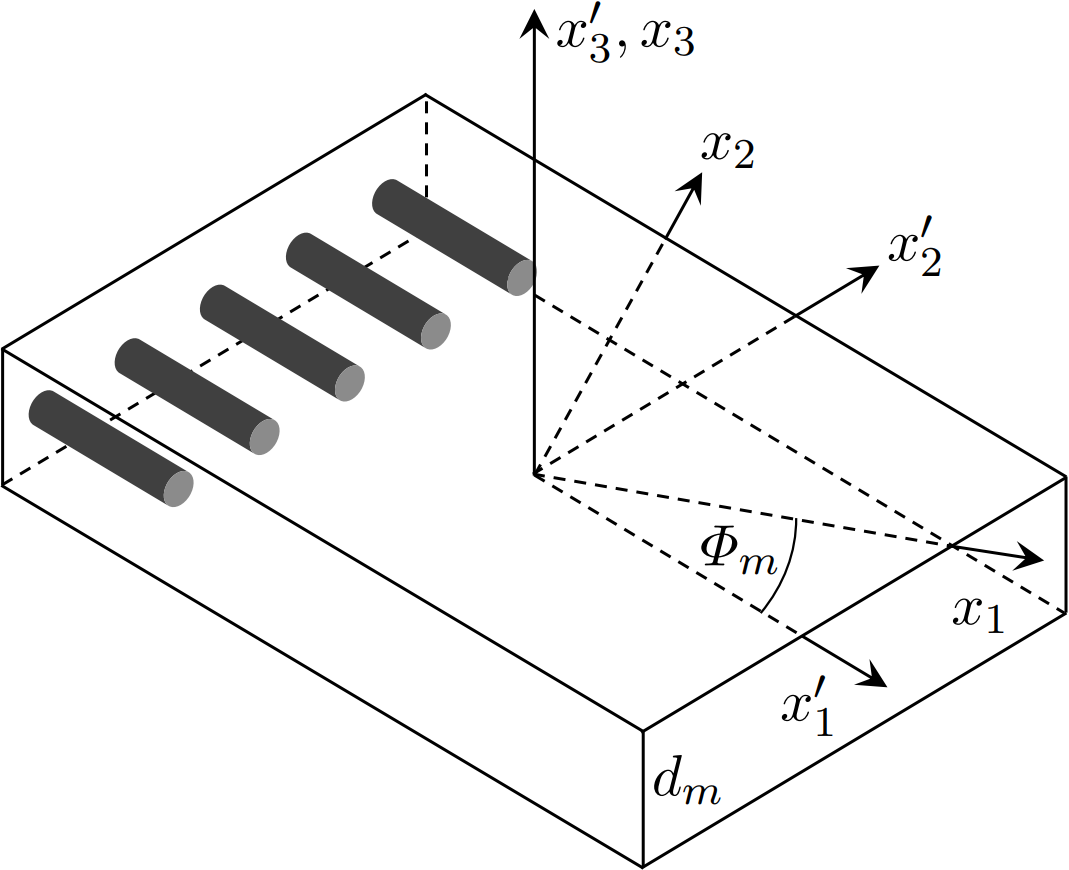}}
\hspace{2mm}
\parbox{.5\textwidth}{
	\includegraphics[width=.47\textwidth]{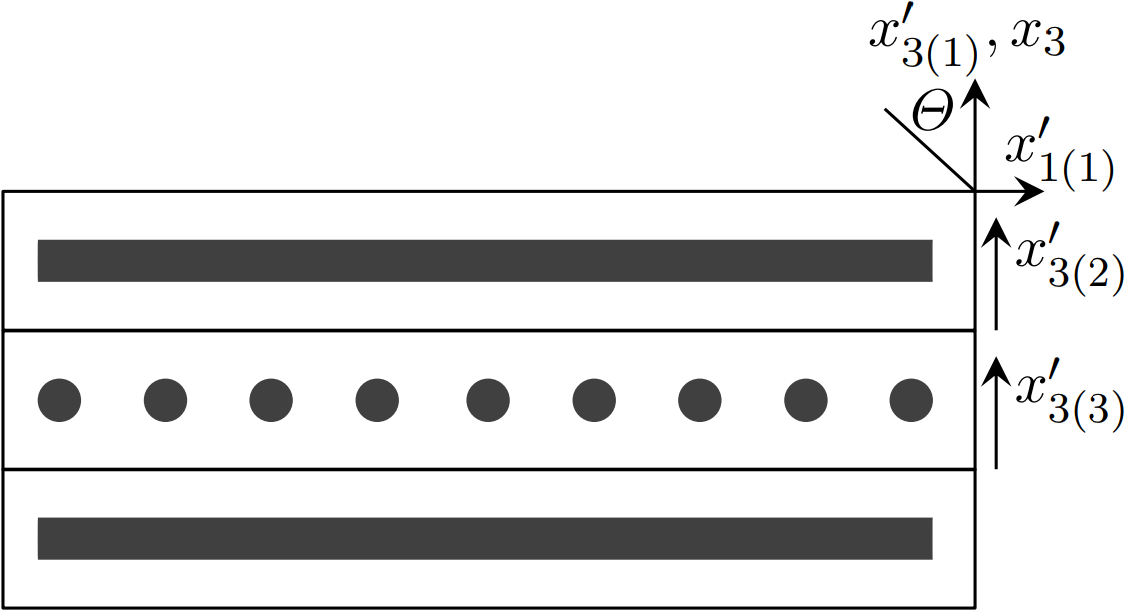}}
\vspace{8mm}			
\caption{\small Schematic of (a) A single composite layer with local (crystallographic) coordinate system $x'_i$ and global coordinate system $x_i$, $i$ = 1, 2, 3. (b) A layered composite plate with [0/90/0] orientation with respect to the $x'_1$-axis.} 
\label{fig.CoordinateSystems}
\end{figure}

Multilayered composites are stacking of \textit{m} layers with layer thicknesses $d_m$ with fiber-epoxy combination in each layer as illustrated in Fig.~\ref{fig.CoordinateSystems}. For each layer, a local (crystallographic) coordinate system $x'_{i(m)}=(x'_1,x'_2,x'_3)_{(m)}$ is assigned at the top of the $m^{th}$ layer, and the layers are defined parallel to the $x'_1$-$x'_2$-plane. The fibers are oriented along $x'_{1(m)}$ direction whereas $x'_{3(m)}$ is normal to the layer. The guided wave propagation in this system with arbitrary layer orientations is described using a global coordinate system $x_i=(x_1,x_2,x_3)$, where guided wave propagation takes place along the $x_1$-direction. With respect to the global coordinate system, the local coordinate systems are yielded by a counterclockwise rotation of an angle $\varPhi_m$ between $x_1$ and $x'_{1(m)}$ about the $x_3$-axis. Here, guided wave propagation is considered in composites containing layers with arbitrary fiber orientations $\varPhi_m$ and with arbitrary propagation angles $\varPhi$ in the $x'_1$-$x'_2$-plane. Correspondingly, the stiffness tensor $c'_{ijkl}$ is transformed from the local to the global coordinate system for each layer and their respective transformed stiffness tensors $c_{ijkl(m)}$ is obtained \cite{dlr139819}.

While solving the forward problem, two assumptions are made on the waveguide. A rigid bonding between the layers is established, implying the continuity of stresses and displacements over the layers' boundaries. It is also assumed that the plate is surrounded by a vacuum, which means that no energy can leak into the surrounding medium. Therefore, the stress components at the top ($x_3=0$) and bottom ($x_3=-d$) of the plate are made to vanish.
%\begin{equation}\label{eq.101}
%u_i=\bar{u}_i,\hspace{5mm}\sigma_{i3}=\bar{\sigma}_{i3},\hspace{5mm}i=1,2,3,
%\end{equation}
%where $u_i$, $\sigma_{i3}$ correspond to the originating layer and $\bar{u}_i$, $\bar{\sigma}_{i3}$ to the continuing layer. 
%Secondly, it is assumed that the plate is surrounded by a vacuum, which means that no energy can leak into the surrounding medium. Therefore, we require the vanishing of the stress components at the top ($x_3=0$) and bottom ($x_3=-d$) of the plate, namely

%\begin{equation}\label{eq.102}
%\sigma^0_{i3},\sigma^{-d}_{i3}=0,\hspace{5mm}i=1,2,3.
%\end{equation}

\noindent The Christoffel equation is formulated by combining Eqs.~(\ref{eq.01}) - (\ref{eq.03}), arriving at Eq.~(\ref{eq.07}).
\begin{equation}\label{eq.07}
\rho\frac{\partial^2 u_{i}}{\partial t^2}=c_{ijkl}\frac{\partial^2 u_l}{\partial x_j\partial x_k}.
\end{equation}

Here, it is intended to solve Christoffel's Eq.~(\ref{eq.07}) for the propagation directions of the bulk waves for which all of them have the same wavenumber component $\xi$ ($=\omega/c_\mathrm p$) along $x_1$, as required by Snell's law. In the general case, the layers of arbitrary fiber orientations and propagation angles, it is assumed that wave motion in the sagittal plane ($x_1$-$x_3$) and shear horizontal motion ($x_2$) are coupled. Therefore, a general form of solution for the displacement field components $u_i$ is set up in terms of the bulk wave amplitudes $U_i$ as shown in Eq.~(\ref{eq.08}).
\begin{equation}\label{eq.08}
(u_1,u_2,u_3)=(U_1,U_2,U_3)\mathrm e^{\mathrm i\xi(x_1+\alpha x_3-c_\mathrm pt)},
\end{equation}
where $c_\mathrm p$ is the phase velocity component along $x_1$ and $\alpha$ is the ratio of the bulk waves' wavenumber components along the $x_3$ and $x_1$-directions ($\alpha=\zeta_3/\zeta_1=\zeta_3/\xi$).

By substituting Eq.~(\ref{eq.08}) into the equation of motion (\ref{eq.02}), three coupled equations are obtained as given by Eq.~(\ref{eq.09}), referred to here as the expanded form of Christoffel's equation, while the $3\times3$ matrix is the corresponding Christoffel matrix $M_{ij}$ where $i,j=1,2,3$. 
\begin{equation}\label{eq.09}
\begin{bmatrix}
	C_{11}-\rho c_\mathrm p^2+C_{55}\alpha^2&C_{16}+C_{45}\alpha^2&(C_{13}+C_{55})\alpha\\
	&C_{66}-\rho c_\mathrm p^2+C_{44}\alpha^2&(C_{36}+C_{45})\alpha\\
	\mathrm{sym}&&C_{55}-\rho c_\mathrm p^2+C_{33}\alpha^2
\end{bmatrix}
\begin{bmatrix}
	U_1\\
	U_2\\
	U_3
\end{bmatrix}=0,
\end{equation}

Three homogeneous linear equations for the displacement amplitudes $U_i$ of the bulk waves are obtained using Eq.~(\ref{eq.09}). Nontrivial solutions for $U_1$, $U_2$, and $U_3$ require the vanishing of the determinant of the Christoffel matrix in Eq.~(\ref{eq.09}), yielding the sixth-degree polynomial equation

\begin{equation}\label{eq.10}
\alpha^6+A_1\alpha^4+A_2\alpha^2+A_3=0,
\end{equation}
with the coefficients $A_1$, $A_2$, and $A_3$ given in Ref.~\cite{dlr139819}. Eq.~(\ref{eq.10}) has six solutions $\alpha_q$, $q=1,2,...,6$. Analytical expressions for the roots are found in Ref.~\cite{dlr139819}. Three pairs of solutions are obtained, corresponding to the downward and upward propagating bulk waves,

\begin{equation}\label{eq.11}
\alpha_{\mathrm{L}^-}=-\alpha_{\mathrm{L}^+},\hspace{5mm}\alpha_{\mathrm{S^-_{fast}}}=-\alpha_{\mathrm{S^+_{fast}}},\hspace{5mm}\alpha_{\mathrm{S^-_{slow}}}=-\alpha_{\mathrm{S^+_{slow}}}.
\end{equation}

Substituting $\alpha_q$ into expanded form of Christoffel's Eq.~(\ref{eq.09}) gives the bulk wave's displacement amplitude ratios $V_q=U_{2q}/U_{1q}$ and $W_q=U_{3q}/U_{1q}$, and with further manipulations gives Eq.~(\ref{eq.12}) \& (\ref{eq.13}).

\begin{equation}\label{eq.12}
V_q=\frac{m_{11}(\alpha_q)m_{23}(\alpha_q)-m_{13}(\alpha_q)m_{12}(\alpha_q)}{m_{13}(\alpha_q)m_{22}(\alpha_q)-m_{12}(\alpha_q)m_{23}(\alpha_q)},
\end{equation}

\begin{equation}\label{eq.13}
W_q=\frac{m_{11}(\alpha_q)m_{22}(\alpha_q)-m_{12}(\alpha_q)^2}{m_{12}(\alpha_q)m_{23}(\alpha_q)-m_{22}(\alpha_q)m_{13}(\alpha_q)},
\end{equation}
where the elements $m_{ij}(\alpha_q)$ are the components of the Christoffel matrix. Now, for convenience, the new variable $\sigma_{ij}^*=\sigma_{ij}/\mathrm i\xi$ is introduced. Thus, the displacement and stress field components are given by Eq.~(\ref{eq.14}).
\begin{equation}\label{eq.14}
\begin{split}
	(u_1,u_2,u_3)&=\sum_{q=1}^{6}(1,V_q,W_q)U_{1q}\mathrm e^{\mathrm i\xi(x_1+\alpha_q x_3-c_\mathrm pt)},\\
	(\sigma_{33}^*,\sigma_{13}^*,\sigma_{23}^*)&=\sum_{q=1}^{6}(D_{1q},D_{2q},D_{3q})U_{1q}\mathrm e^{\mathrm i\xi(x_1+\alpha_q x_3-c_\mathrm pt)},
\end{split}
\end{equation}
where the stress amplitudes are obtained from the Hooke's law (Eq.~\ref{eq.03}) as shown in Eq.~(\ref{eq.15}).
\begin{equation}\label{eq.15}
\begin{split}
	D_{1q}&=C_{13}+C_{36}V_q+C_{33}\alpha_q W_q,\\
	D_{2q}&=C_{55}(\alpha_q+W_q)+C_{45}\alpha_q V_q,\\
	D_{3q}&=C_{45}(\alpha_q+W_q)+C_{44}\alpha_q V_q.
\end{split}
\end{equation}

Eq.~(\ref{eq.14}) can be simplified using the symmetry property of $\alpha_q$ as presented in Eq.~(\ref{eq.11}). The amplitude ratios in Eqs.~(\ref{eq.12}) and (\ref{eq.13}) have the properties $V_2 = V_1$, $V_4 = V_3$, $V_6 = V_5$, $W_2 = -W_1$, $W_4 =-W_3$, and $W_6 = -W_5$. Likewise the stress amplitudes have the properties $D_{12} = D_{11}$, $D_{14} = D_{13}$, $D_{16} = D_{15} $, $D_{22} = -D_{21}$, $D_{24} = -D_{23}$, $D_{26} = -D_{25}$, $D_{32} = -D_{31}$, $D_{34} = -D_{33}$, and $D_{36} = -D_{35}$.

Eq.~(\ref{eq.14}) can be rewritten in matrix form as shown in Eq.~(\ref{eq.16}), relating the displacement and stress components at the top $\textit{\textbf{u}}_m$, $\boldsymbol\sigma_{m}^*$ ($x_{3(m)}=0$) and bottom $\textit{\textbf{u}}_{m+1}$, $\boldsymbol\sigma_{m+1}^*$ ($x_{3(m)}=-d_m$) of the $m^{th}$ layer to the wave amplitude vectors $\textit{\textbf{U}}_m^\pm$.

\begin{equation}\label{eq.16}
\begin{split}
	\begin{bmatrix}
		\textit{\textbf{u}}_m\\
		\textit{\textbf{u}}_{m+1}
	\end{bmatrix}&=
	\begin{bmatrix}
		\textit{\textbf{P}}^-&\textit{\textbf{P}}^+\textit{\textbf{H}}\\
		\textit{\textbf{P}}^-\textit{\textbf{H}}&\textit{\textbf{P}}^+
	\end{bmatrix}_m
	\begin{bmatrix}
		\textit{\textbf{U}}_m^-\\
		\textit{\textbf{U}}_m^+
	\end{bmatrix},\\
	\begin{bmatrix}
		\boldsymbol\sigma_{m}^*\\
		\boldsymbol\sigma_{m+1}^*
	\end{bmatrix}&=
	\begin{bmatrix}
		\textit{\textbf{D}}^-&\textit{\textbf{D}}^+\textit{\textbf{H}}\\
		\textit{\textbf{D}}^-\textit{\textbf{H}}&\textit{\textbf{D}}^+
	\end{bmatrix}_m
	\begin{bmatrix}
		\textit{\textbf{U}}_m^-\\
		\textit{\textbf{U}}_m^+
	\end{bmatrix},
\end{split}
\end{equation}
where $d_m$ is the thickness of the $m^{th}$ layer and  
\begin{align}\label{eq.16a}
\textit{\textbf{P}}^-&=\begin{bmatrix}
	1&1&1\\
	V_1&V_3&V_5\\
	W_1&W_3&W_5   
\end{bmatrix},   &   \textit{\textbf{P}}^+&=\begin{bmatrix}
	1&1&1\\
	V_1&V_3&V_5\\
	-W_1&-W_3&-W_5   
\end{bmatrix}\nonumber,\\
\textit{\textbf{D}}^-&=\begin{bmatrix}
	D_{11}&D_{13}&D_{15}\\
	D_{21}&D_{23}&D_{25}\\
	D_{31}&D_{33}&D_{35}   
\end{bmatrix},   &   \textit{\textbf{D}}^+&=\begin{bmatrix}
	D_{11}&D_{13}&D_{15}\\
	-D_{21}&-D_{23}&-D_{25}\\
	-D_{31}&-D_{33}&-D_{35}  
\end{bmatrix},\\
\textit{\textbf{H}}&=\begin{bmatrix}
	\mathrm{e}^{\mathrm{i}\xi\alpha_1 x_3} &0&0\\
	0&\mathrm{e}^{\mathrm{i}\xi\alpha_3 x_3} &0\\
	0&0&\mathrm{e}^{\mathrm{i}\xi\alpha_5 x_3}   
\end{bmatrix}\nonumber,   &   \textit{\textbf{U}}^\pm&=\begin{bmatrix}
	U_{11}\\
	U_{12}\\
	U_{13}  
\end{bmatrix},
\end{align}
where the common factor $\mathrm{e}^{\mathrm{i}\xi(x_1-c_\mathrm pt)}$ is suppressed for brevity. By eliminating the displacement amplitude vectors $\textit{\textbf{U}}_m^\pm$ from Eq.~(\ref{eq.16}), the stresses at the top $\boldsymbol\sigma_m^*$ and bottom $\boldsymbol\sigma_{m+1}^*$ of the $m^{th}$ layer are related to the displacements at the top $\textit{\textbf{u}}_m$ and bottom $\textit{\textbf{u}}_{m+1}$ via the stiffness matrix $\textit{\textbf{K}}_m$ as given by Eq.~(\ref{eq.17}).

\begin{equation}\label{eq.17}
\begin{bmatrix}
	\boldsymbol\sigma_m^*\\
	\boldsymbol\sigma_{m+1}^*
\end{bmatrix}=
\begin{bmatrix}
	\textit{\textbf{D}}^-&\textit{\textbf{D}}^+\textit{\textbf{H}}\\
	\textit{\textbf{D}}^-\textit{\textbf{H}}&\textit{\textbf{D}}^+
\end{bmatrix}_m
\begin{bmatrix}
	\textit{\textbf{P}}^-&\textit{\textbf{P}}^+\textit{\textbf{H}}\\
	\textit{\textbf{P}}^-\textit{\textbf{H}}&\textit{\textbf{P}}^+
\end{bmatrix}_m^{-1}
\begin{bmatrix}
	\textit{\textbf{u}}_m\\
	\textit{\textbf{u}}_{m+1}
\end{bmatrix}
=\textit{\textbf{K}}_m
\begin{bmatrix}
	\textit{\textbf{u}}_m\\
	\textit{\textbf{u}}_{m+1}
\end{bmatrix}.
\end{equation}

In order to obtain the global stiffness matrix of a multilayered plate, first, the local layer stiffness matrices $\textit{\textbf{K}}_m$ is calculated. For instance, the local stiffness matrices of two neighboring layers (1-2 and 2-3) is shown by Eq.~(\ref{eq.18}).
\begin{equation}\label{eq.18}
\begin{bmatrix}
	\boldsymbol\sigma_1^*\\
	\boldsymbol\sigma_2^*
\end{bmatrix}=
\begin{bmatrix}
	\textit{\textbf{K}}_{11}^A&\textit{\textbf{K}}_{12}^A\\
	\textit{\textbf{K}}_{21}^A&\textit{\textbf{K}}_{22}^A
\end{bmatrix}
\begin{bmatrix}
	\textit{\textbf{u}}_1\\
	\textit{\textbf{u}}_2
\end{bmatrix},\hspace{5mm}
\begin{bmatrix}
	\boldsymbol\sigma_2^*\\
	\boldsymbol\sigma_3^*
\end{bmatrix}=
\begin{bmatrix}
	\textit{\textbf{K}}_{11}^B&\textit{\textbf{K}}_{12}^B\\
	\textit{\textbf{K}}_{21}^B&\textit{\textbf{K}}_{22}^B
\end{bmatrix}
\begin{bmatrix}
	\textit{\textbf{u}}_2\\
	\textit{\textbf{u}}_3
\end{bmatrix}.
\end{equation}

Rokhlin and Wang's recursive algorithm given by Eq.~(\ref{eq.19}) is used to combine both.
\begin{equation}\label{eq.19}
\begin{bmatrix}
	\boldsymbol\sigma_1^*\\
	\boldsymbol\sigma_3^*
\end{bmatrix}=
\begin{bmatrix}
	\textit{\textbf{K}}_{11}^A+\textit{\textbf{K}}_{12}^A(\textit{\textbf{K}}_{11}^B-\textit{\textbf{K}}_{22}^A)^{-1}\textit{\textbf{K}}_{21}^A&-\textit{\textbf{K}}_{12}^A(\textit{\textbf{K}}_{11}^B-\textit{\textbf{K}}_{22}^A)^{-1}\textit{\textbf{K}}_{12}^B\\
	\textit{\textbf{K}}_{21}^B(\textit{\textbf{K}}_{11}^B-\textit{\textbf{K}}_{22}^A)^{-1}\textit{\textbf{K}}_{21}^A&\textit{\textbf{K}}_{22}^B-\textit{\textbf{K}}_{21}^B(\textit{\textbf{K}}_{11}^B-\textit{\textbf{K}}_{22}^A)^{-1}\textit{\textbf{K}}_{12}^B
\end{bmatrix}
\begin{bmatrix}
	\textit{\textbf{u}}_1\\
	\textit{\textbf{u}}_3
\end{bmatrix}.
\end{equation}

Calling the obtained matrix as $\textit{\textbf{K}}^A$ and the stiffness matrix of the next layer i.e., third layer as $\textit{\textbf{K}}^B$, Eq.~(\ref{eq.19}) can be used recursively to obtain the global stiffness matrix $\textit{\textbf{K}}$, relating the stresses and displacements at the top and bottom of the whole plate. To find modal solutions, the stress components at the top ($x_3=0$) and bottom ($x_3=-d$) of the plate are made to vanish. This leads to the characteristic equation~(\ref{eq.20}). The SMM is discussed in Ref.~\cite{huber2018classification,dlr139819} in more detail. 
\begin{equation}\label{eq.20}
\mathrm{det}\textit{\textbf{K}}=0.
\end{equation}

The group velocity $c_{\mathrm g}$ is considered in above formulations but energy velocity $c_{\mathrm e}$ is the more general term  \cite{giurgiutiu2021stress}. The energy velocity is equal to the group velocity in non-dissipative systems, i.e., without attenuation caused by viscoelasticity or energy leakage to the surrounding medium. The group velocity direction (ray angle $\varPhi_\mathrm r$) deviates by the skew angle $\gamma$ from the wave propagation direction $x_1$ (wave propagation angle $\varPhi$)
\begin{equation}\label{eq.06a}
\varPhi_\mathrm r=\varPhi-\gamma.
\end{equation}
The group velocity direction is described by a vector with a component $c_{\mathrm g1}$ aligned in the wave propagation direction $x_1$, and a component $c_{\mathrm g2}$ normal to it ($x_2$). Hence, the group velocity magnitude $c_\mathrm g$ is given by
\begin{equation}\label{eq.07a}
c_\mathrm g=c_\mathrm e=|\vec{c}_\mathrm g|=\sqrt{c_{\mathrm g1}^2+c_{\mathrm g2}^2}.
\end{equation}
$\vec{c}_\mathrm g$ points into the direction of energy flow, which indicates the actual direction of the guided wave beam. The skew angle is calculated using Eq.~(\ref{eq.08a}).
\begin{equation}\label{eq.08a}
\gamma=-\mathrm{tan}^{-1}\frac{c_{\mathrm g2}}{c_{\mathrm g1}}
\end{equation}
The negative sign in Eq.~(\ref{eq.08a}) accounts for the fact that the guided wave beam, i.e., the energy flow, skews towards the fiber direction because energy can be transported more efficiently in that direction. The group velocity components $c_{\mathrm g1}$ and $c_{\mathrm g2}$ are calculated from the ratios of the power flow $P_j$ and the total energy $E_\mathrm{tot}$ using Eq.~(\ref{eq.09a}).
\begin{equation}\label{eq.09a}
\vec{c}_\mathrm g=\vec{c}_\mathrm e=\frac{1}{E_\mathrm{tot}}
\begin{bmatrix}
	P_1\\
	P_2\\
	0
\end{bmatrix}
=
\begin{bmatrix}
	c_{\mathrm g1}\\
	c_{\mathrm g2}\\
	0
\end{bmatrix}
\end{equation}
The total energy is the energy per unit volume carried by a guided wave. It has two contributions, namely the strain energy $E_{\mathrm{strain}}$ and the kinetic energy $E_{\mathrm{kin}}$ such that $E_{\mathrm{tot}}=E_{\mathrm{strain}}+E_{\mathrm{kin}}$, where 
\begin{equation}\label{eq.01a}
E_{\mathrm{strain}}=\frac{1}{2}\int_{S}\sigma_{ij}\varepsilon_{ij}\mathrm dS,\hspace{5mm}i,j=1,2,3,
\end{equation}
with the stress tensor $\sigma_{ij}$ and the strain tensor $\varepsilon_{ij}$, and
\begin{equation}\label{eq.02a}
E_{\mathrm{kin}}=\frac{1}{2}\rho\int_{S}v_i^2\mathrm dS,
\end{equation}
where integration of the respective energy densities is performed over the cross-section $S$ of the laminate, and where $\rho$ is the material's density and $v_i=\dot{u_i}$ is the particle velocity vector. The power flow $P_j$ is the energy flow per unit volume and unit time carried by a guided wave. It is obtained using Eq.~(\ref{eq.03a}) by integrating the power flow density (\textit{Poynting vector}) $p_j$, indicating the magnitude and direction of the power flow. 
\begin{equation}\label{eq.03a}
P_j=\int_{S}p_j\mathrm dS=-\frac{1}{2}\int_{S}\mathrm{Re}(\sigma_{ij}v^*_i)\mathrm dS=-\frac{1}{2}\int_{S}\mathrm{Re}
\begin{bmatrix}
	\sigma_{11}v^*_1+\sigma_{21}v^*_2+\sigma_{31}v^*_3\\
	\sigma_{12}v^*_1+\sigma_{22}v^*_2+\sigma_{32}v^*_3\\
	\sigma_{13}v^*_1+\sigma_{23}v^*_2+\sigma_{33}v^*_3
\end{bmatrix}\mathrm dS,
\end{equation}
where the star ($^*$) indicates the complex conjugate. Combining Eqs.~(\ref{eq.01a}) - (\ref{eq.03a}) yields
\begin{equation}\label{eq.10a}
\begin{split}
	E_{\mathrm{tot}}=\frac{1}{2}\int_{S}(\sigma_{11}\varepsilon_{11}+\sigma_{33}\varepsilon_{33}+\sigma_{23}\varepsilon_{23}+\sigma_{13}\varepsilon_{13}+\sigma_{12}\varepsilon_{12})\mathrm dS,\\
	+\frac{1}{2}\rho\int_{S} (v_1^2+v_2^2+v_3^2)\mathrm dS,
\end{split}
\end{equation}
($\varepsilon_{22}$ is zero), and the power flow components are calculated as
\begin{equation}\label{eq.11a}
\begin{split}
	P_1=-\frac{1}{2}\int_{S}\mathrm{Re}(\sigma_{11}v^*_1+\sigma_{12}v^*_2+\sigma_{13}v^*_3)\mathrm dS,\\		P_2=-\frac{1}{2}\int_{S}\mathrm{Re}(\sigma_{12}v^*_1+\sigma_{22}v^*_2+\sigma_{23}v^*_3)\mathrm dS.
\end{split}
\end{equation}

\begin{figure}[h!]
%\captionsetup{justification=centering}
\centering
\begin{minipage}[b]{0.45\textwidth}
	\centering
	\includegraphics[width=0.8\textwidth]{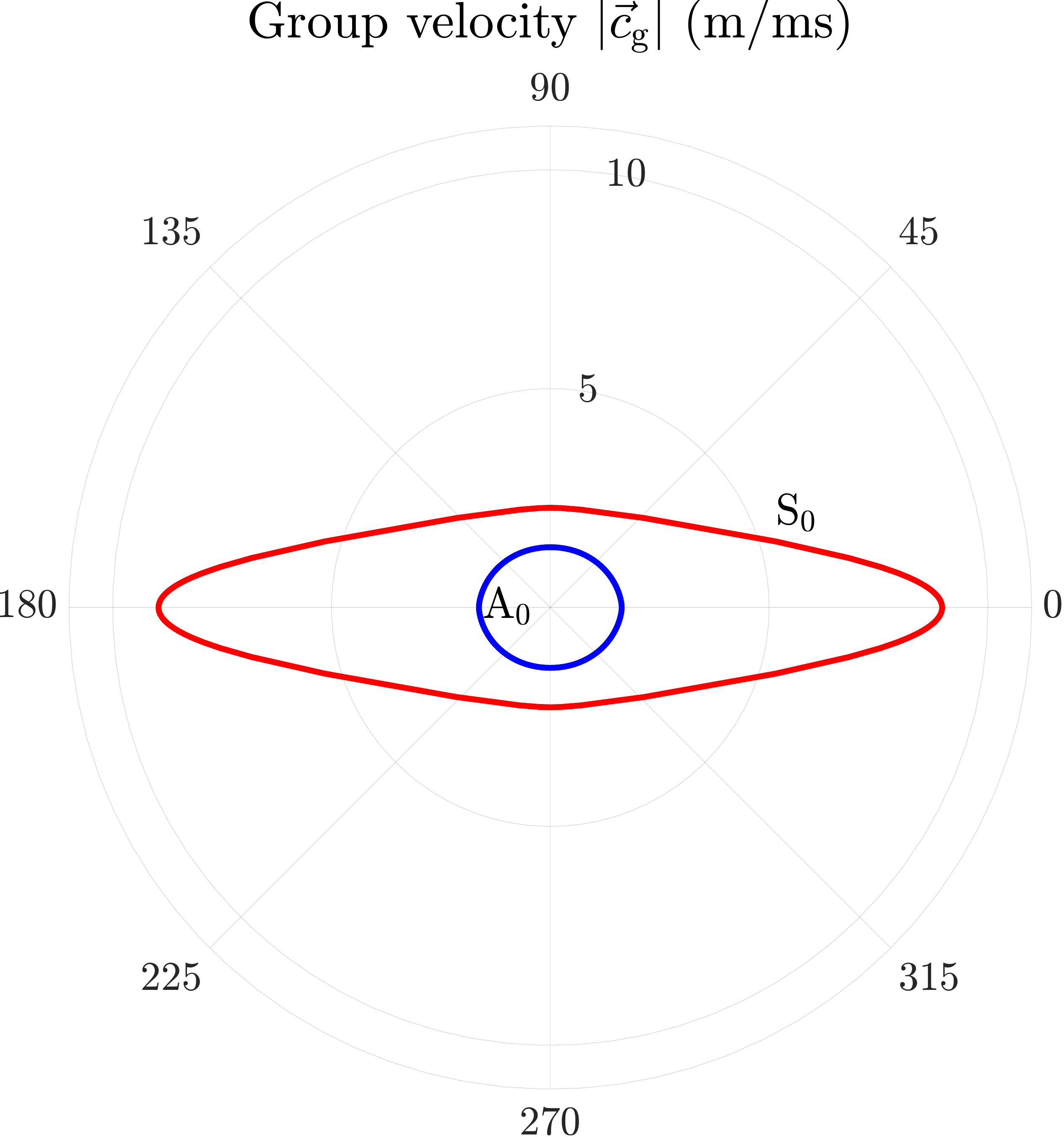}
	%\vspace{1ex}
\end{minipage}
\begin{minipage}[b]{0.45\textwidth}
	\centering
	\includegraphics[width=0.8\textwidth]{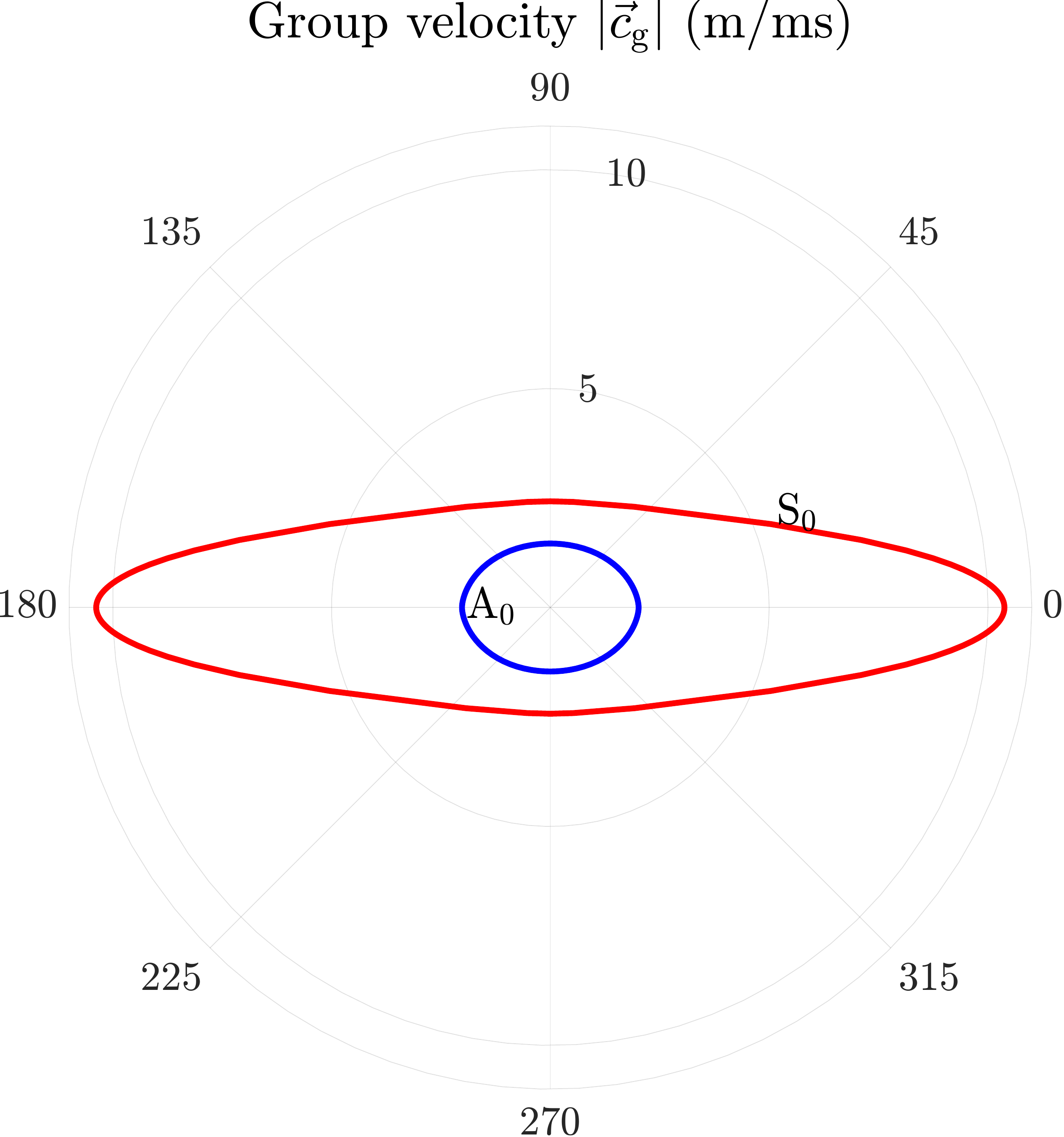}
	%\vspace{4ex}
\end{minipage}
\caption{\small Group velocity magnitude $|\vec{c}_{\mathrm g}|$ of the fundamental modes A$_0$ and S$_0$ in 2\,mm thick unidirectional composite layers (a) T700M21 (b) T800M924 at 200\,kHz.}
\label{fig:PolarDiagram}
\end{figure}

The power flow component normal to the specimen, $P_3$ is zero because there is no net energy transport across the thickness of the laminate. Fig.~\ref{fig:PolarDiagram} shows the group velocity magnitudes $|\vec{c}_{\mathrm g}|$ of the fundamental modes A$_0$ and S$_0$ versus the ray angle $\varPhi_\mathrm r$ for a full sweep of the wave propagation angle $\varPhi=[0\,^\circ \hspace{2mm} 360\,^\circ]$ in a 2\,mm thick unidirectional composite layers for two commercially available Carbon Fiber Reinforced Polymer (CFRP) materials i.e., T700M21 (Fig.~\ref{fig:PolarDiagram}(a)) and T800M924 (Fig.~\ref{fig:PolarDiagram}(b)) at a frequency of 200\,kHz (The properties of other commercially available CFRP composite materials are given in Table-\ref{tab:ComMat}). These plots, also called wavecrest plots, show how the guided waves generated by a point source propagate outwards with time. Fig.~\ref{fig:PolarDiagram} illustrates how much the fiber direction ($\varPhi=0\,^\circ$) is preferred by S$_0$ for the transport of energy. In this direction, the composite has the highest stiffness, which leads to maximum group (and phase) velocities. It is calculated as 8.950\,m/ms in T700M21 and 10.374\,m/ms in T800M924.

\begin{table}[h!]
\centering
\captionsetup{justification=centering}
\setlength{\belowcaptionskip}{0pt}
\caption{\small Material properties of commercially available CFRP composite materials}
\addtolength{\tabcolsep}{-1pt} 
\begin{tabular}{l|c c c c c c}
	\hline
	Commerical CFRP & $\rho$  & $E_1$ & $E_2$ & $G_{12}$ & $\nu_{12}$ & $\nu_{23}$\\
	composite materials &(kg/$\mathrm{m}^3$) & (GPa) & (GPa) & (GPa)& &\\
	\cline{1-7}
	AS4M3502 \cite{SAUSE2018291} & 1550 & 144.6 & 9.6 & 6 & 0.3 & 0.28 \\
	GraphiteEpoxy\_Rokhlin\_2011 \cite{rokhlin2011physical} & 1610 & 150.95 & 12.8 & 8 & 0.47 & 0.45 \\
	SAERTEX7006919RIMR135 \cite{huber2018dispersion} & 1454 & 119.9 & 7.25 & 6 & 0.32 & 0.45 \\
	SigrafilCE125023039 \cite{SAUSE2018291} & 1500 & 128.6 & 6.87 & 6.1 & 0.33 & 0.37 \\
	T300M914 \cite{moll2019open} & 1560 & 139.92 & 10.05 & 5.7 & 0.31 & 0.48\\
	T700M21 \cite{Simon1997}  & 1571 & 125.5 & 8.7 & 4.1 & 0.37 & 0.45\\
	T700PPS \cite{SAUSE2018291} & 1600 & 149.96 & 9.98 & 4.5 & 0.29 & 0.37 \\
	T800M913 \cite{SAUSE2018291} & 1550 & 152.14 & 6.64 & 20 & 0.25 & 0.54 \\
	T800M924 \cite{percival1997study} & 1500 & 161 & 9.25  & 6 & 0.34 & 0.41\\
	T800\_Michel \cite{yu2017feature} & 1510 & 178.96 & 9.17 & 5.5 & 0.36 & 0.53\\
	\hline
\end{tabular}
\label{tab:ComMat}
\end{table}

\begin{figure}[h!]
\centering
\includegraphics[width=1.0\textwidth]{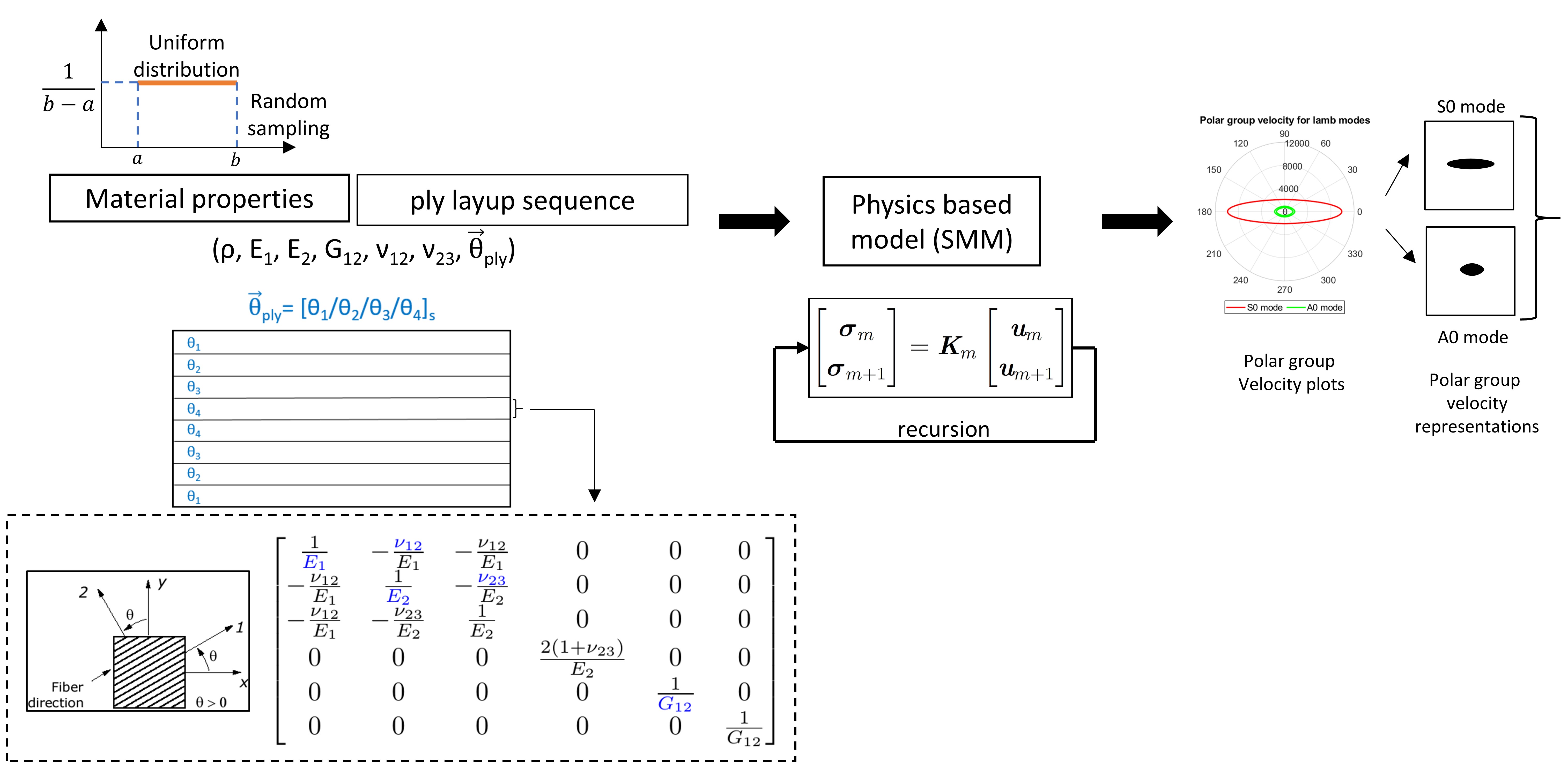}
\caption{\small Detailed schematic of the forward model: Material properties and ply-layup sequence are the inputs to the model and polar representations are the outputs.}
\label{fig:forward}	
\end{figure}

The entire approach of using SMM and the group velocity calculation routine as the forward model is presented schematically in Fig.~\ref{fig:forward}. In this procedure, either commercial CFRP materials (Table-\ref{tab:ComMat}) are used, or the material properties are randomly sampled from a uniform distribution based on the range of each material property present in the commercial CFRP composites. A vector with six elements for each of the three layup sequences (unidirectional, cross-ply, quasi-isotropic), representing six material properties (density, Young's modulus in 1 and 2 directions, shear modulus in 1-2 plane, Poisson's ratio in 1-2 and 2-3 plane) is entered into the physics-based model (SMM). Results in the form of polar group velocity at different frequencies are collected from the model. Polar group velocities are transformed into binary images (black \& white) named as polar representations to exploit the image processing capabilities of CNN. The dataset collected from the forward model is used to train the inverse model.

%%%%

\subsection{Inverse model: Dual-branch CNN}\label{ssec:inverse}
\begin{figure}[h!]
\centering
\includegraphics[width=1.0\textwidth]{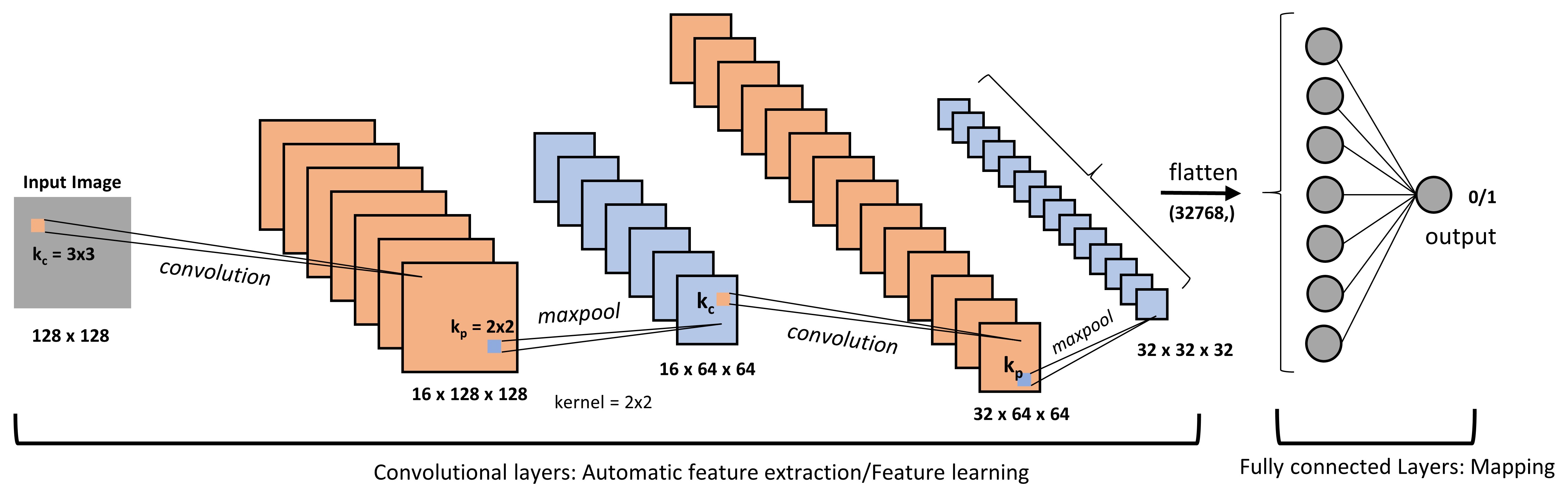}
\caption{\small A schematic of a general convolutional neural networks having two convolutional layers and one fully connected layer followed by binary classification output.}
\label{fig:CNN}	
\end{figure}

Neural networks are called a universal function approximator which maps input space (2d representations of polar group velocities) to target space (material properties or layup sequence type) \cite{hornik1989multilayer,barron1993universal}. It is implemented by propagating information in the forward direction to get some output that may not be the true output. A cost function along with an optimization scheme is used to update the parameters (weight and biases) while minimizing the cost function \cite{rautela2019electromechanical}. A deeper neural network has the capability to learn complex and more sophisticated patterns underlying the data \cite{telgarsky2016benefits}. It is carried out by incorporating more composite functions for the mapping. CNN is widely used to process data having grid-like topology or spatial sequences (images) and temporal sequences (time-series data). It has sparse connections and parameter sharing, which makes it distinct from a fully connected network (FCN) \cite{rautela2021ultrasonic,RAUTELA2021106451}. A CNN is constructed using convolutional and pooling layers, which helps it to extract useful features automatically at different levels of abstractions, as shown in Fig.~\ref{fig:CNN}.

\begin{figure}[h!]
\centering
\includegraphics[width=1.0\textwidth]{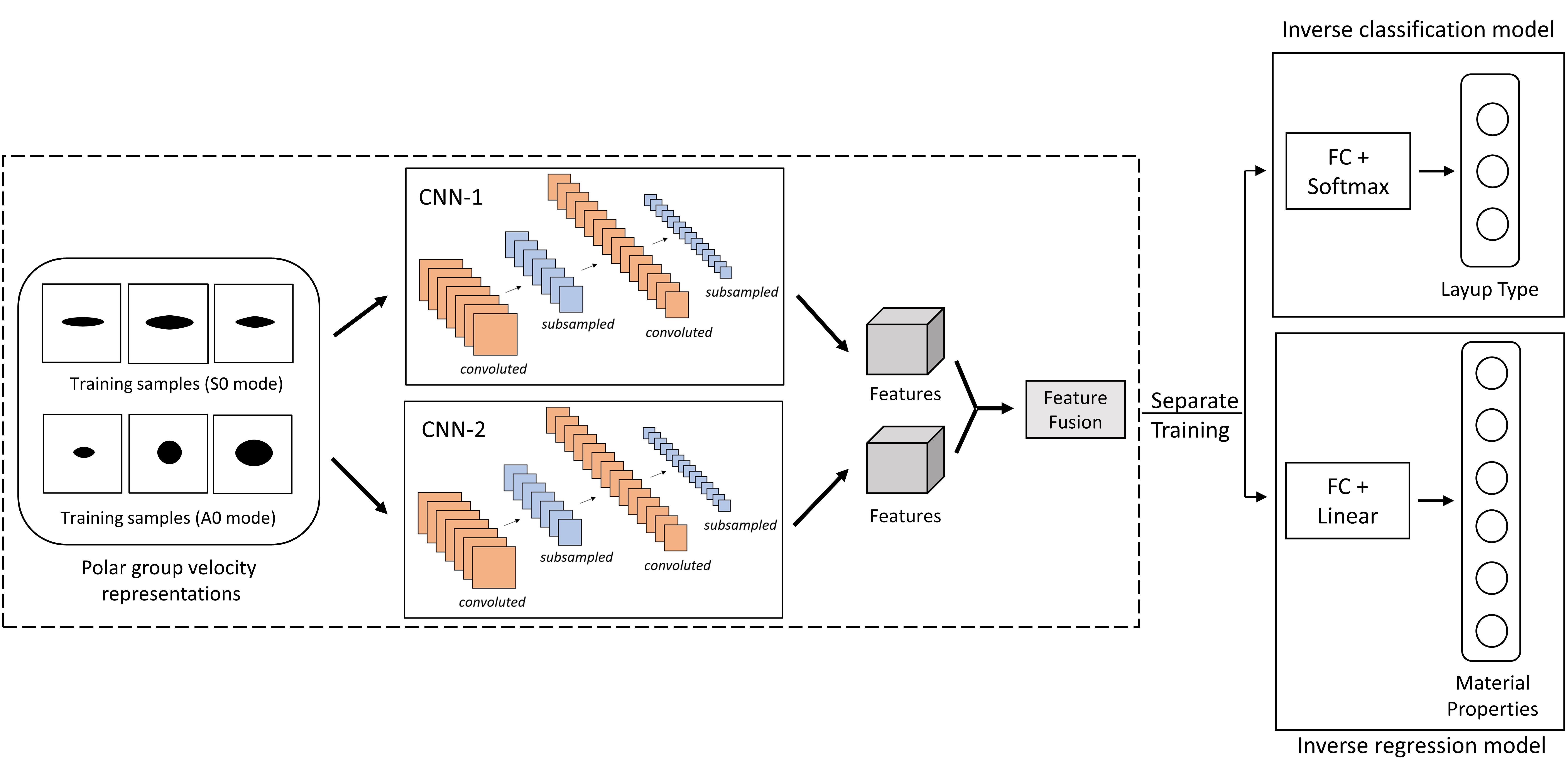}
\caption{\small Schematic of the inverse models (Dual-branch CNN): inputs are polar representations and outputs are ply layup type (classification model) or material properties (regression model).}
\label{fig:inverse}	
\end{figure}

Traditional CNN is used as a single sequential model to extract features hierarchically and subsequently as the image passes through the network. In order to process information coming from independent representations present in the dataset (like different modes of guided waves), different sequential models can be stacked together in the form of parallel branches. Each branch may consist of a CNN to extract features and all features coming from different branches can be flattened and fused together before passing into a FCN \cite{zhao2019multi}. In this study, a dual-branch CNN is used as an inverse model to process multi-modal polar group velocity representations of the guided waves. A classification based dual-branch CNN is utilized to categorize ply layup sequences into three classes and regression-based dual-branch CNN to obtain the corresponding material properties as shown in Fig.~\ref{fig:inverse}.

%---------------------- Begin NEW SECTION
\section{Sensitivity Analysis}\label{sec:sensitivity}
\begin{figure}
\begin{minipage}[b]{1.0\textwidth}
	\centering
	\begin{minipage}[b]{0.3\textwidth}
		\centering
		\includegraphics[width=1.0\textwidth]{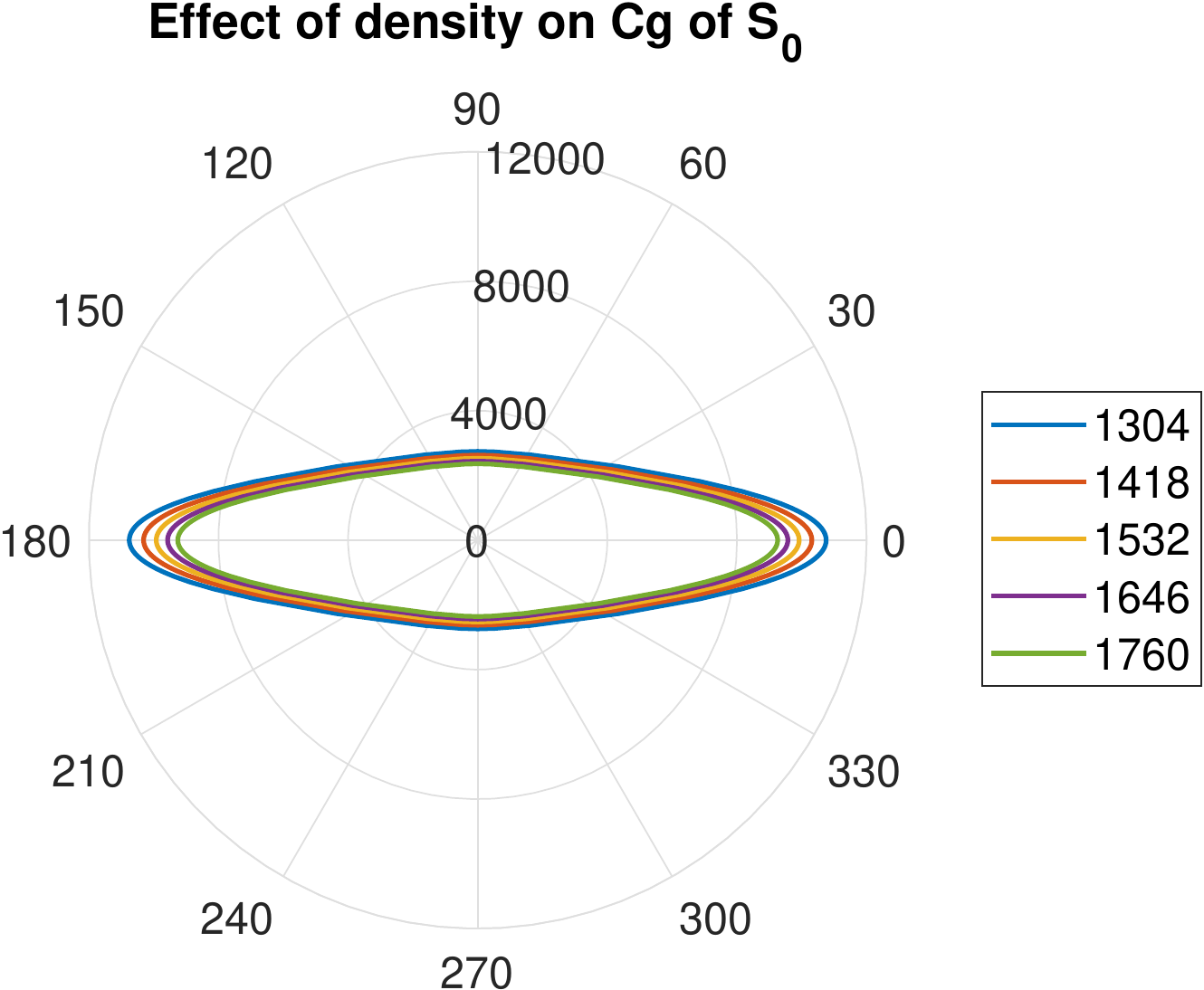}
		%\subcaption{Density effect (kg/$m^3$) on S0}
		\vspace{1ex}
	\end{minipage}
	\begin{minipage}[b]{0.3\textwidth}
		\centering
		\includegraphics[width=1.0\textwidth]{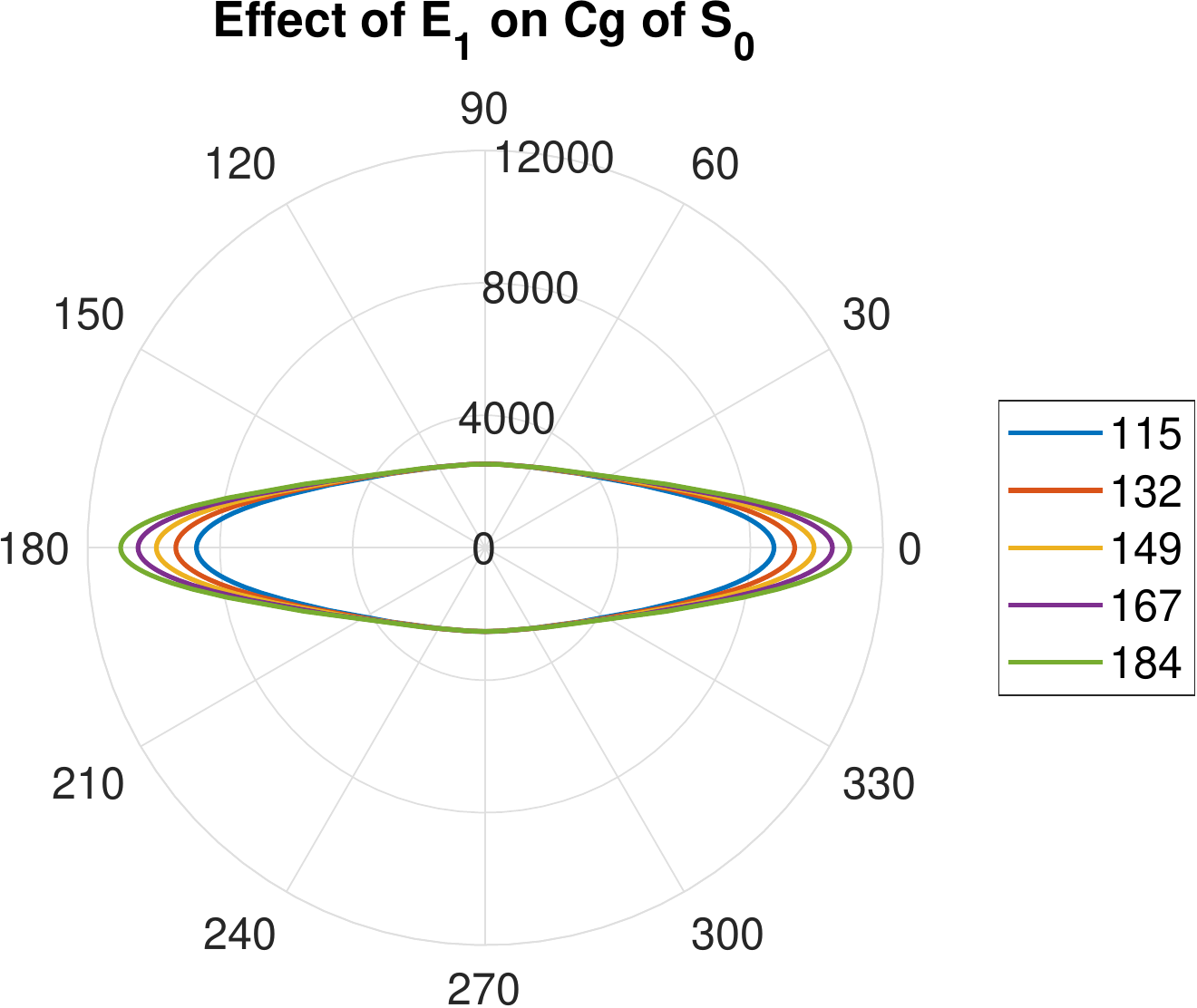}
		%\subcaption{Density effect (kg/$m^3$) on A0}
		\vspace{1ex}
	\end{minipage}
	\begin{minipage}[b]{0.3\textwidth}
		\centering
		\includegraphics[width=1.0\textwidth]{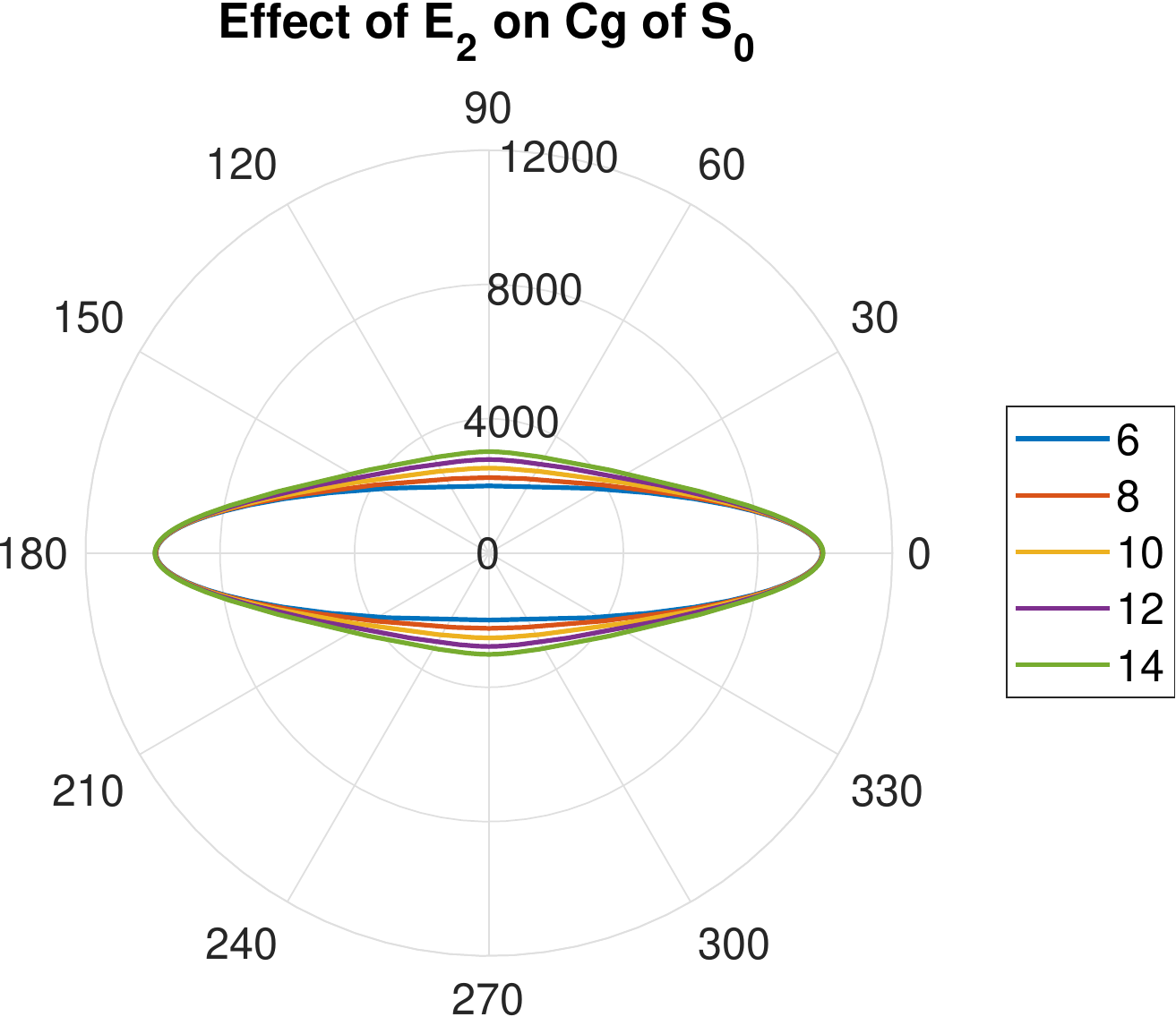}
		%\subcaption{Density effect (kg/$m^3$) on SH0}
		\vspace{1ex}
	\end{minipage}
	\centering
	\begin{minipage}[b]{0.3\textwidth}
		\centering
		\includegraphics[width=1.0\textwidth]{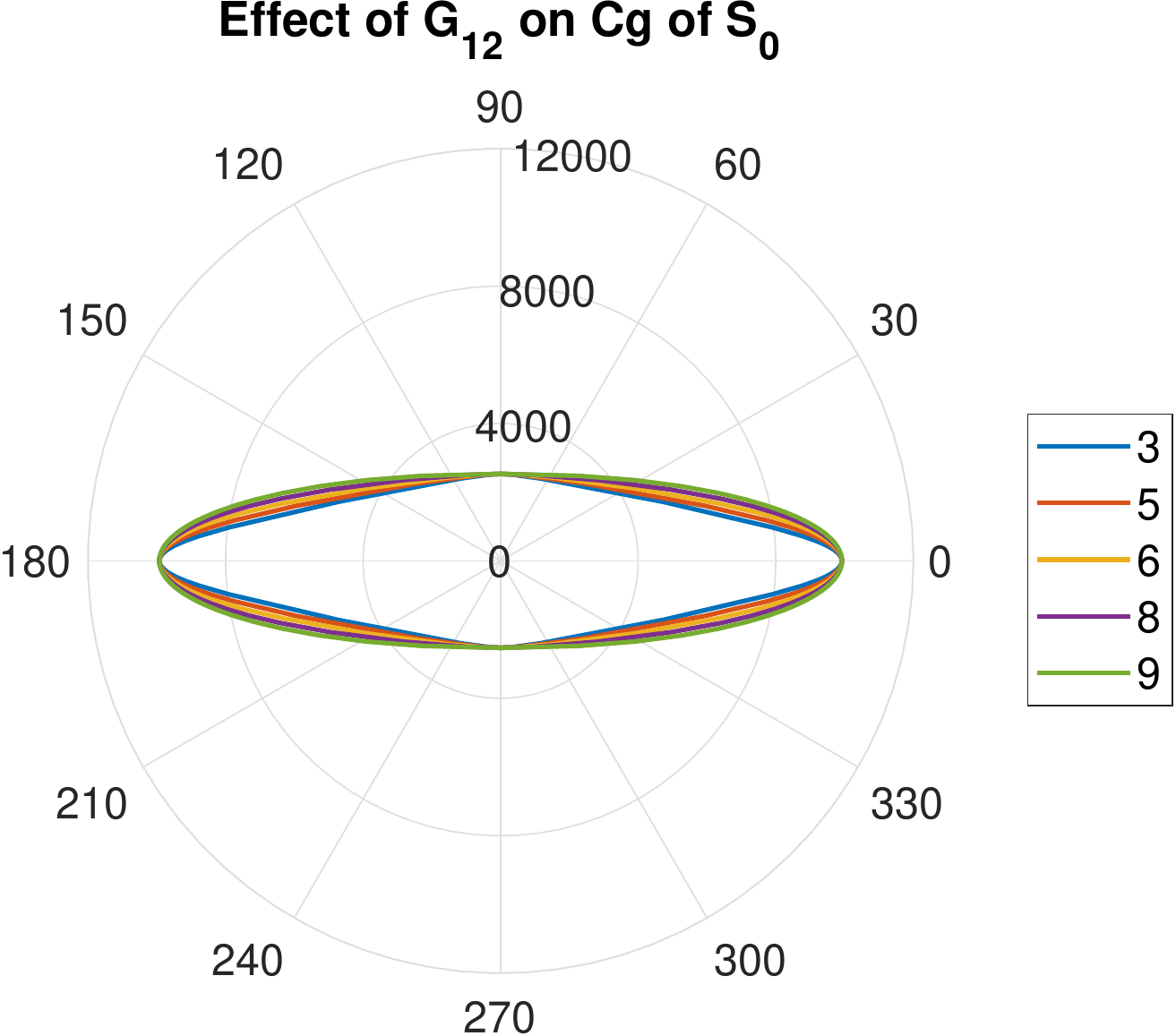}
		%\subcaption{$E_{2}$ effect (GPa) on S0}
		%\vspace{1ex}
	\end{minipage}
	\begin{minipage}[b]{0.3\textwidth}
		\centering
		\includegraphics[width=1.0\textwidth]{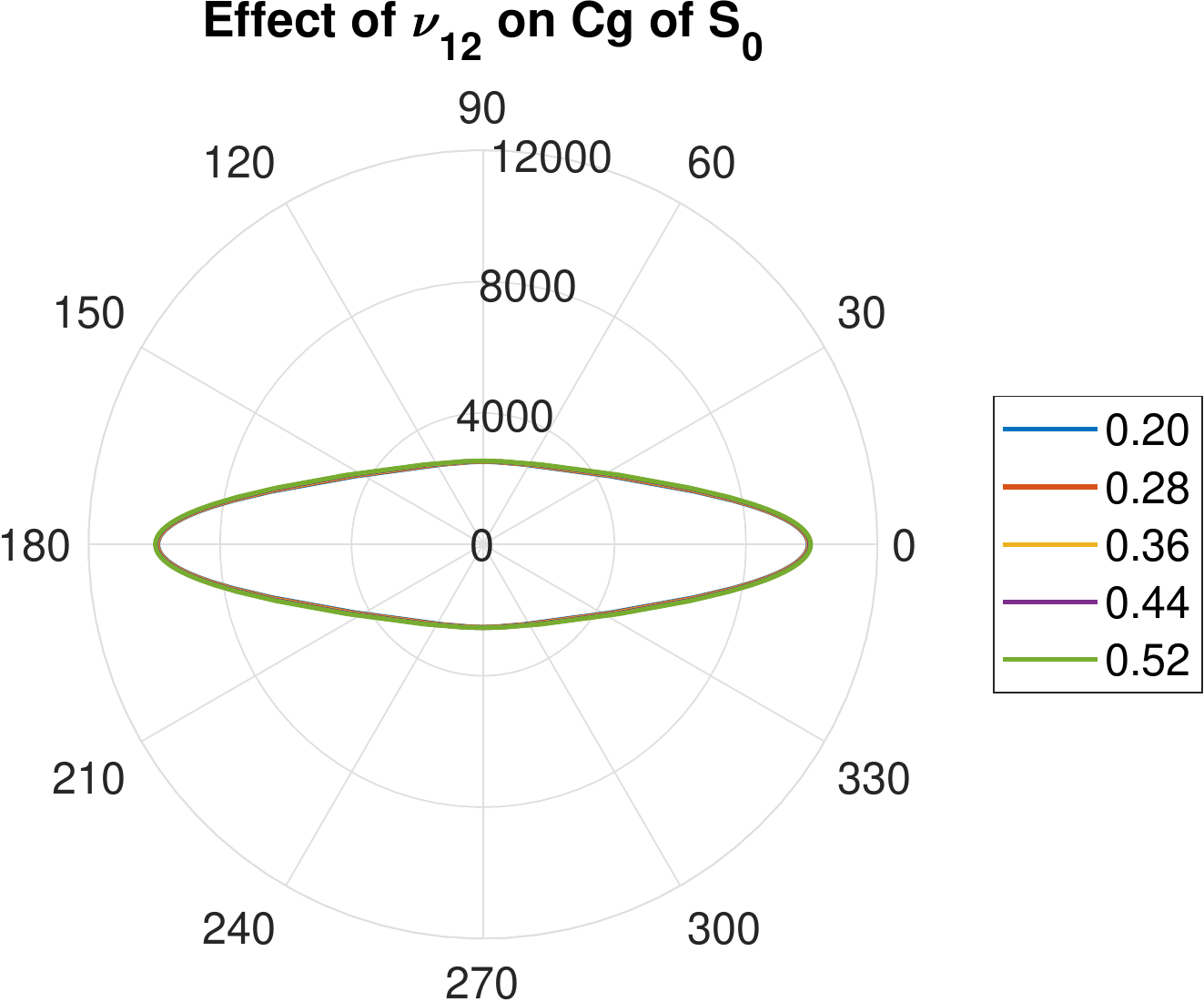}
		%\subcaption{$E_{2}$ effect (GPa) on A0}
		%\vspace{1ex}
	\end{minipage}
	\begin{minipage}[b]{0.3\textwidth}
		\centering
		\includegraphics[width=1.0\textwidth]{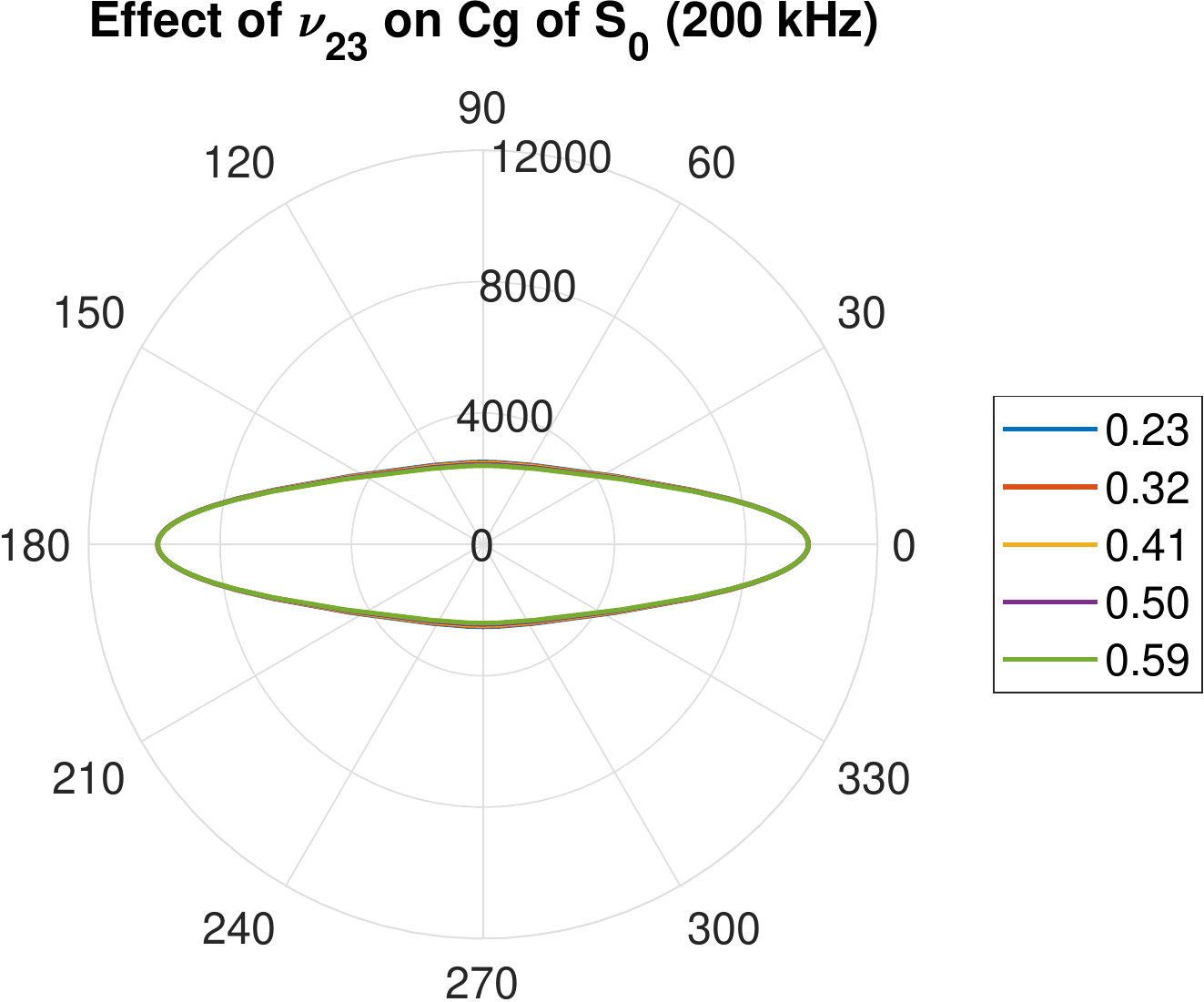}
		%\subcaption{$E_{2}$ effect (GPa) on SH0}
		%\vspace{1ex}
	\end{minipage}
	\subcaption{Effect of material properties on polar group velocity of $S_0$ mode.}
	\vspace{1ex}
\end{minipage}
%%%%
\begin{minipage}{1.0\textwidth}
	\centering
	\begin{minipage}[b]{0.3\textwidth}
		\centering
		\includegraphics[width=1.0\textwidth]{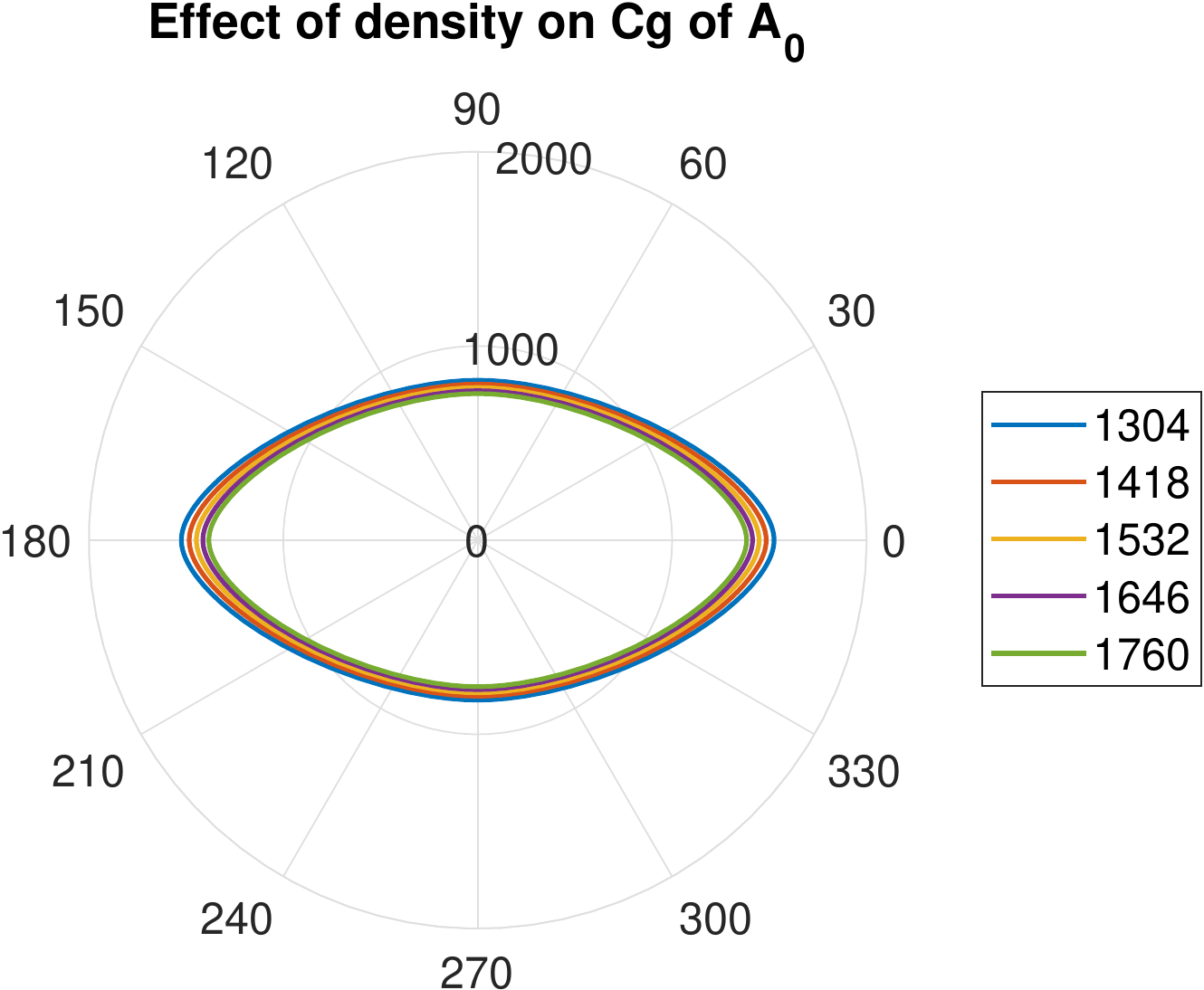}
		%\subcaption{Density effect (kg/$m^3$) on S0}
		\vspace{1ex}
	\end{minipage}
	\begin{minipage}[b]{0.3\textwidth}
		\centering
		\includegraphics[width=1.0\textwidth]{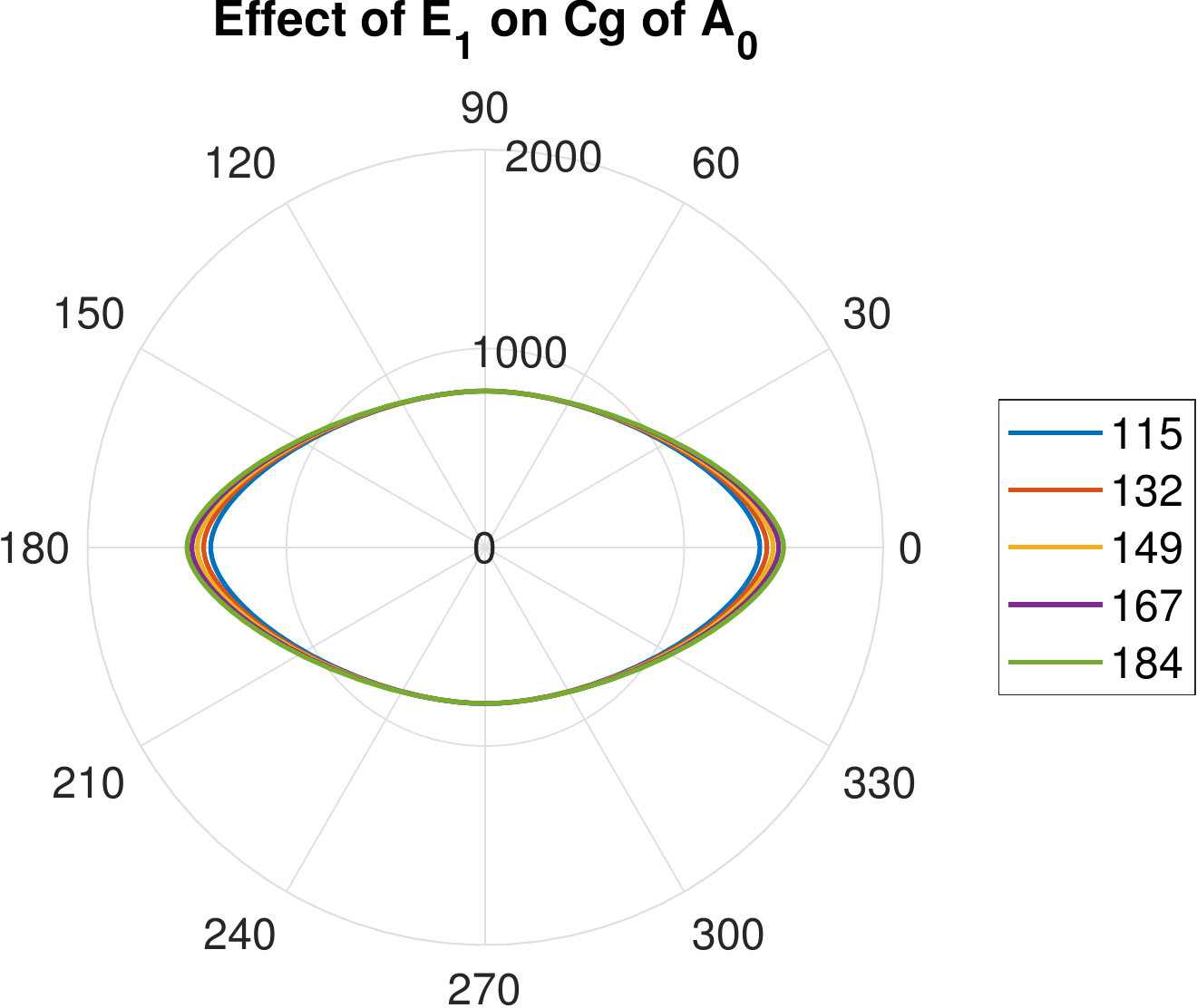}
		%\subcaption{Density effect (kg/$m^3$) on A0}
		\vspace{1ex}
	\end{minipage}
	\begin{minipage}[b]{0.3\textwidth}
		\centering
		\includegraphics[width=1.0\textwidth]{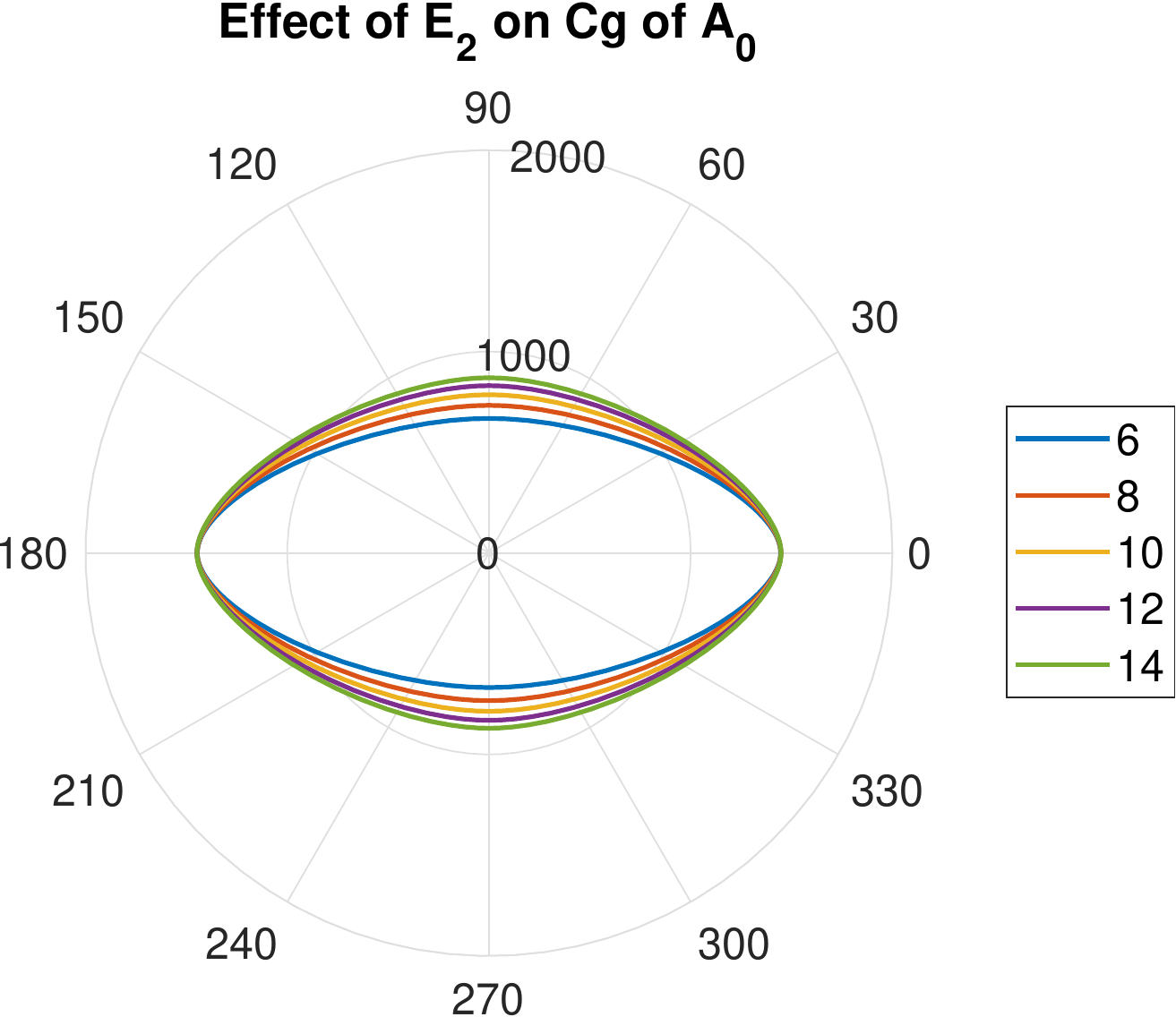}
		%\subcaption{Density effect (kg/$m^3$) on SH0}
		\vspace{1ex}
	\end{minipage}
	\centering
	\begin{minipage}[b]{0.3\textwidth}
		\centering
		\includegraphics[width=1.0\textwidth]{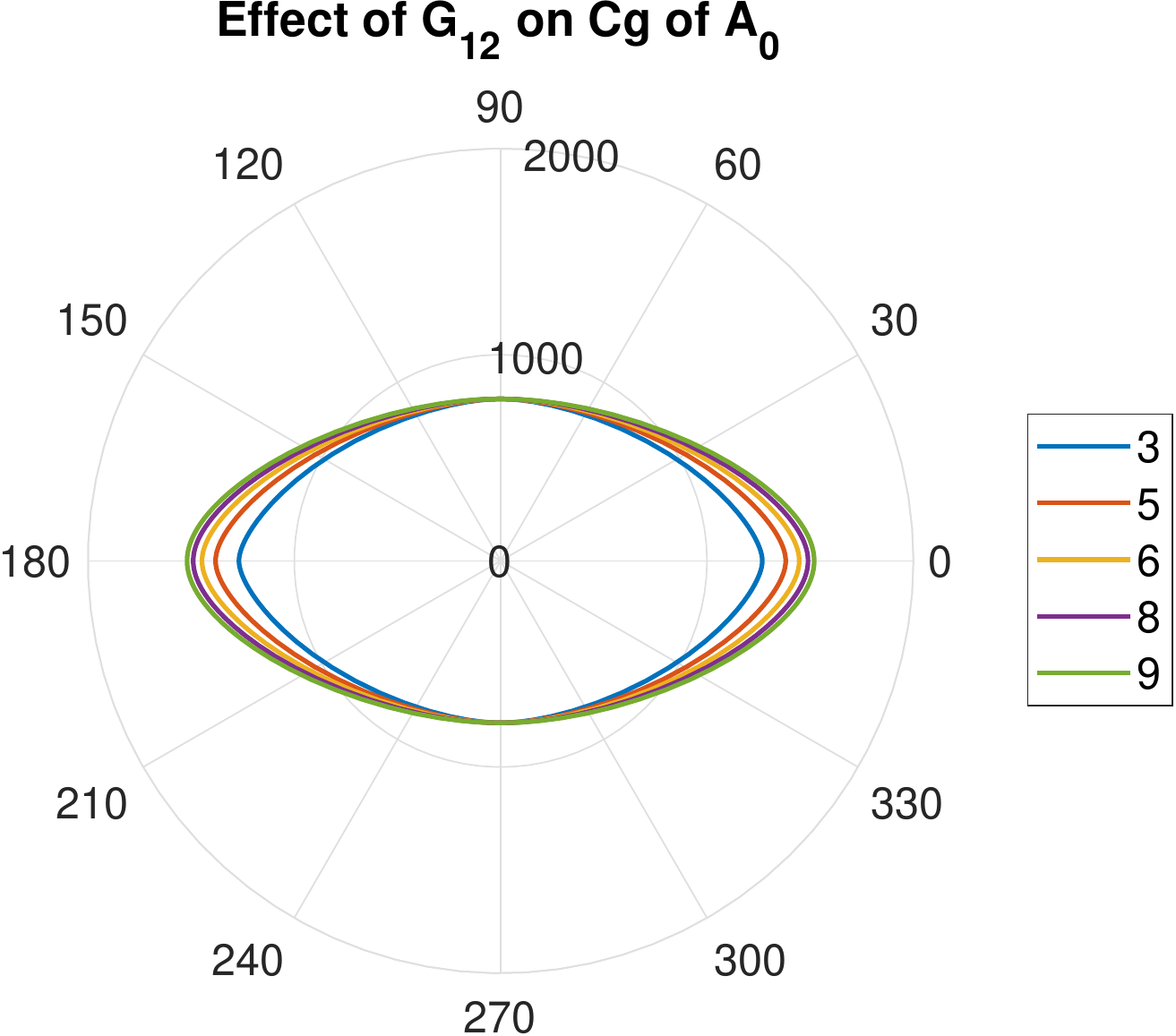}
		%\subcaption{$E_{2}$ effect (GPa) on S0}
		%\vspace{1ex}
	\end{minipage}
	\begin{minipage}[b]{0.3\textwidth}
		\centering
		\includegraphics[width=1.0\textwidth]{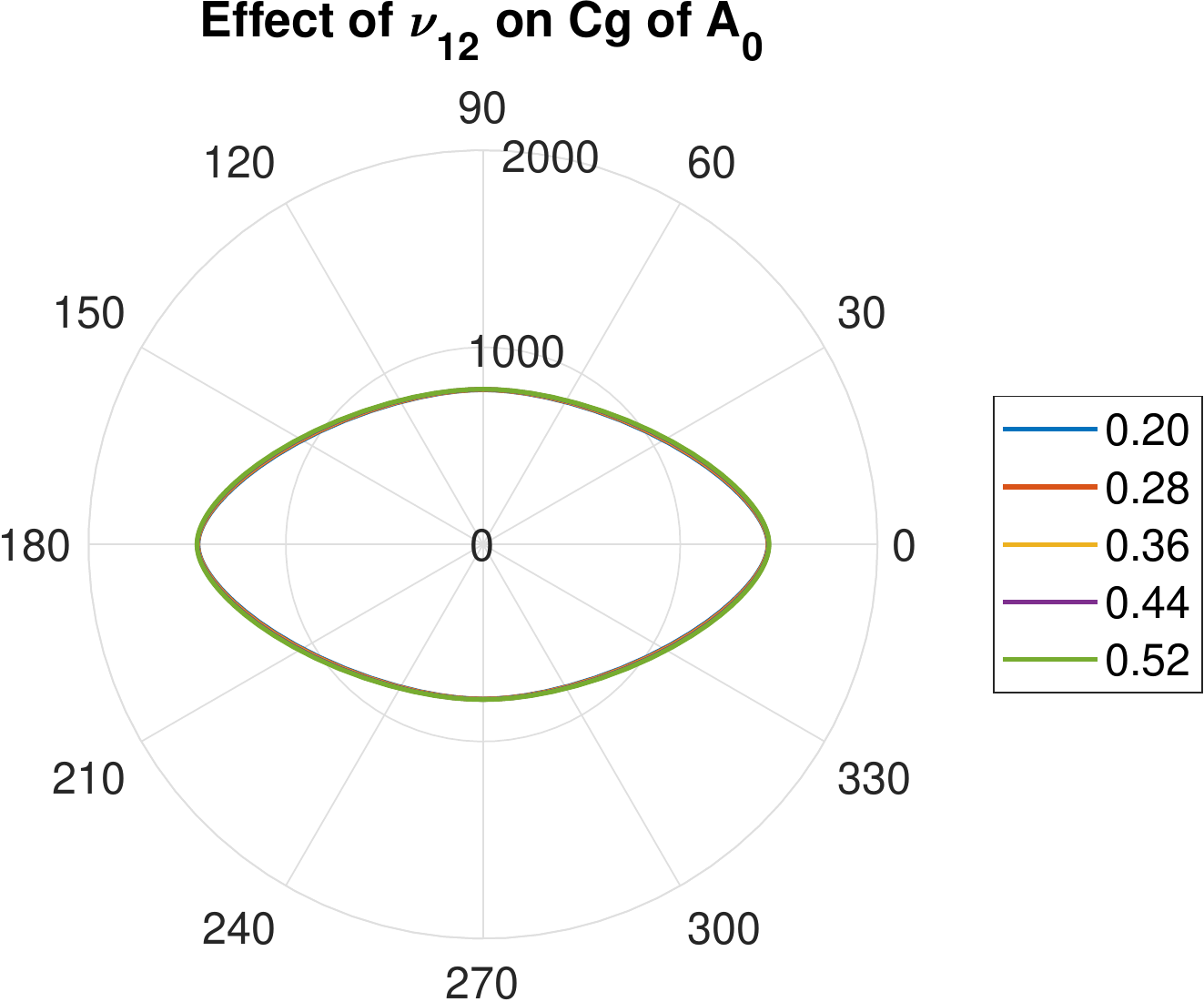}
		%\subcaption{$E_{2}$ effect (GPa) on A0}
		%\vspace{1ex}
	\end{minipage}
	\begin{minipage}[b]{0.3\textwidth}
		\centering
		\includegraphics[width=1.0\textwidth]{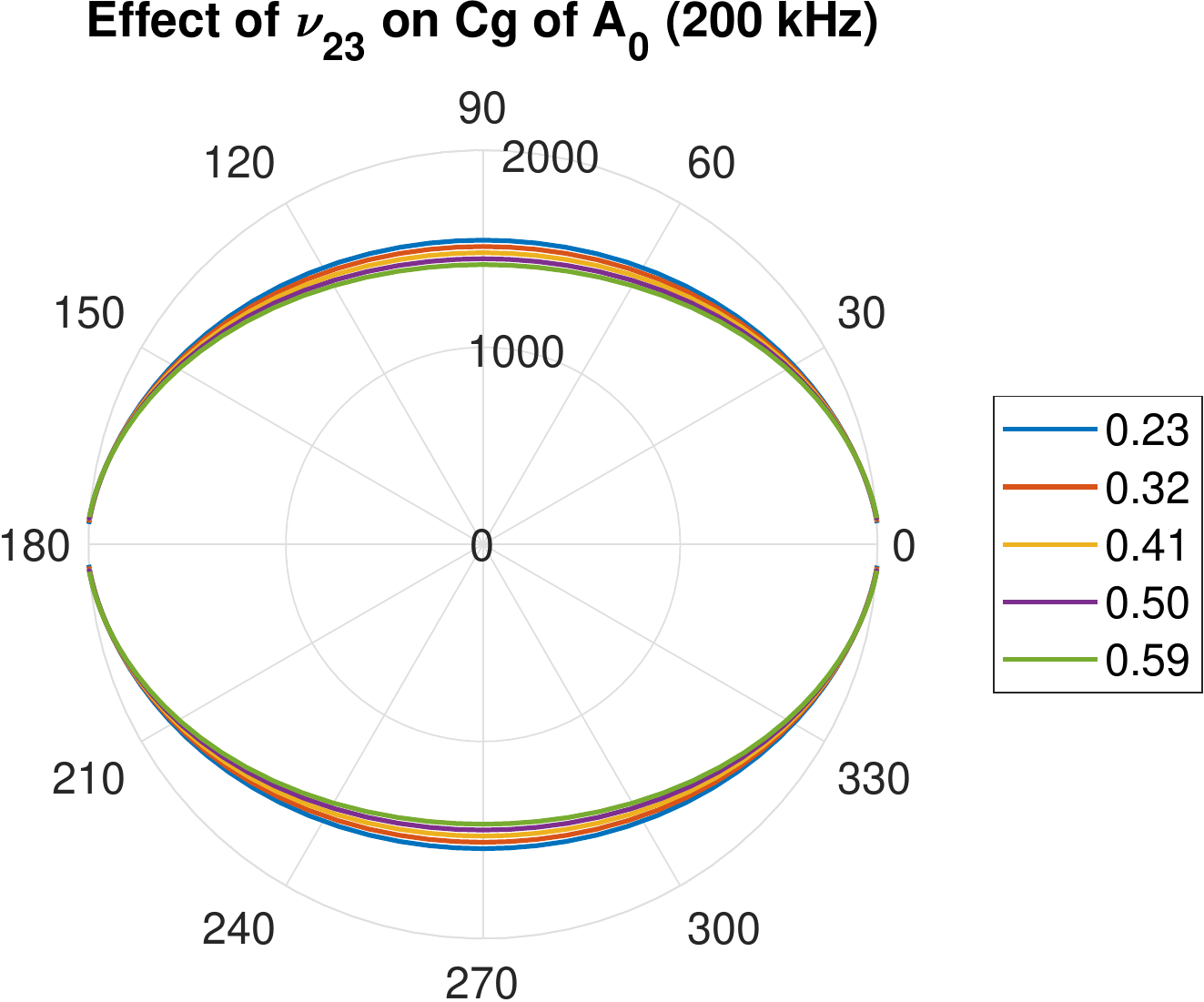}
		%\subcaption{$E_{2}$ effect (GPa) on SH0}
		%\vspace{1ex}
	\end{minipage}
	\subcaption{Effect of material properties on polar group velocity of $A_0$ mode.}
\end{minipage}
\caption{\small Sensitivity analysis: Effect of different material properties i.e.,  $\rho$ (kg/$\mathrm{m}^3$), $E_1$ (GPa), $E_2$ (GPa), $G_{12}$ (GPa), $\nu_{12}$, $\nu_{23}$ on polar group velocities of two fundamental Lamb modes (a) $S_0$ modes (b) $A_0$ mode.}
\label{fig:sensivitiy}
\end{figure}

Sensitivity analysis plays an essential role in the inverse problem of material characterization \cite{balasubramaniam1998inversion,vishnuvardhan2007genetic,cui2019identification}. It helps to understand the effect of inputs on the output, which may assist in the training procedure of the networks for the inverse problem. For a well-posed inverse problem, the output of the forward model (group velocities of two modes) should be influenced in some ways by the permutations and combinations of corresponding inputs (material properties). Here, the effect of density and five elastic constants ($E_1$, $E_2$, $G_{12}$, $\nu_{12}$, $\nu_{23}$) on the polar group velocity of the two Lamb modes (fundamental symmetric and antisymmetric, called $S_0$ and $A_0$, respectively) is studied. Each of the six material properties is varied in a range based on the range of each material property present in commercially available materials (See Table-\ref{tab:ComMat}). In the table, the density range is [1454 1610] kg/$\mathrm{m}^3$, and in order to randomly sample new materials, this range is pushed on both sides by 150 kg/$\mathrm{m}^3$. Therefore, the bounds for the uniform distribution for density become [1304 1760] kg/$\mathrm{m}^3$. Similar exercise is performed for the other material properties, and their respective bounds become [115 184] GPa, [6 14] GPa, [3 9] GPa, [0.20 0.52] and [0.23 0.59] for $E_1$, $E_2$, $G_{12}$, $\nu_{12}$, $\nu_{23}$, respectively. For sensitivity analysis, each material property is varied in this bound while keeping the other properties at their mean values \cite{rautela2020ultrasonicIEEE}. For all such variations, polar group velocities of $S_0$ and $A_0$ modes are plotted using the forward model. The effect of these six parameters on the group velocity of both modes is shown in Fig.~\ref{fig:sensivitiy}.

% Combined FIGURE
% End figure

\begin{figure}
\centering
\begin{minipage}[b]{0.32\textwidth}
	\centering
	\includegraphics[width=1.0\textwidth]{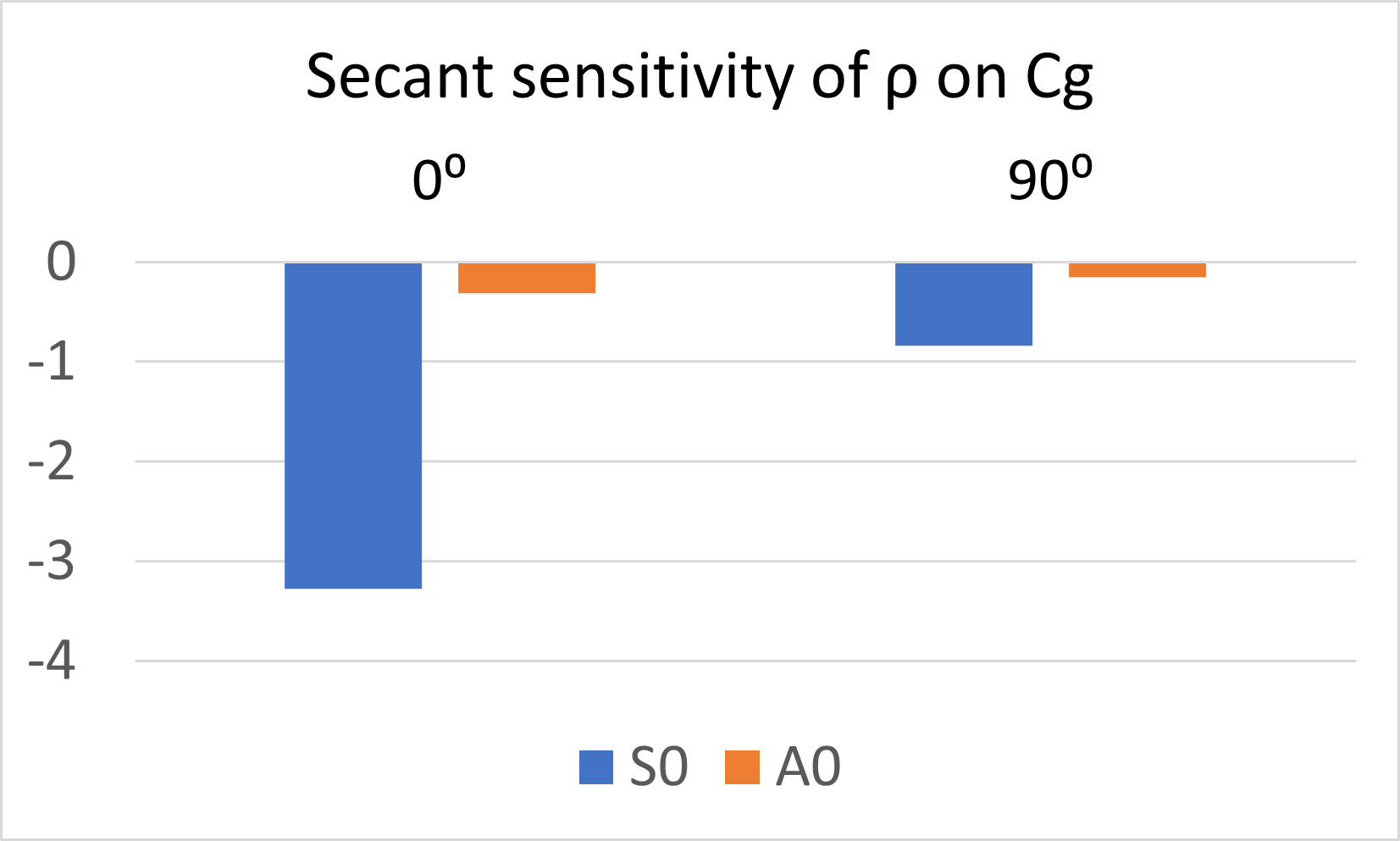}
	%\subcaption{$v_{12}$ effect on $S_0$}
	\vspace{1ex}
\end{minipage}
\begin{minipage}[b]{0.32\textwidth}
	\centering
	\includegraphics[width=1.0\textwidth]{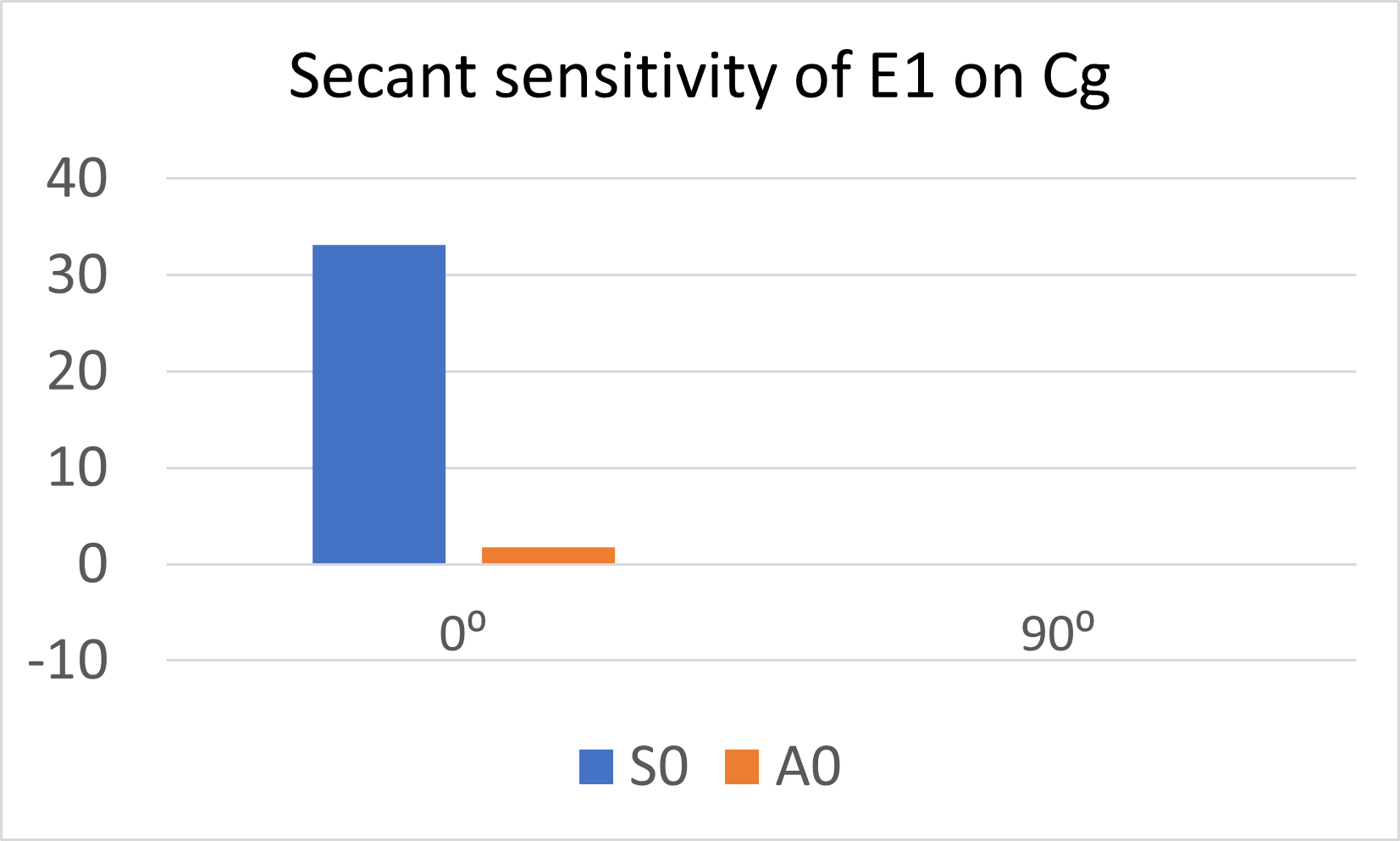}
	%\subcaption{$v_{12}$ effect on A0}
	\vspace{1ex}
\end{minipage}
\begin{minipage}[b]{0.32\textwidth}
	\centering
	\includegraphics[width=1.0\textwidth]{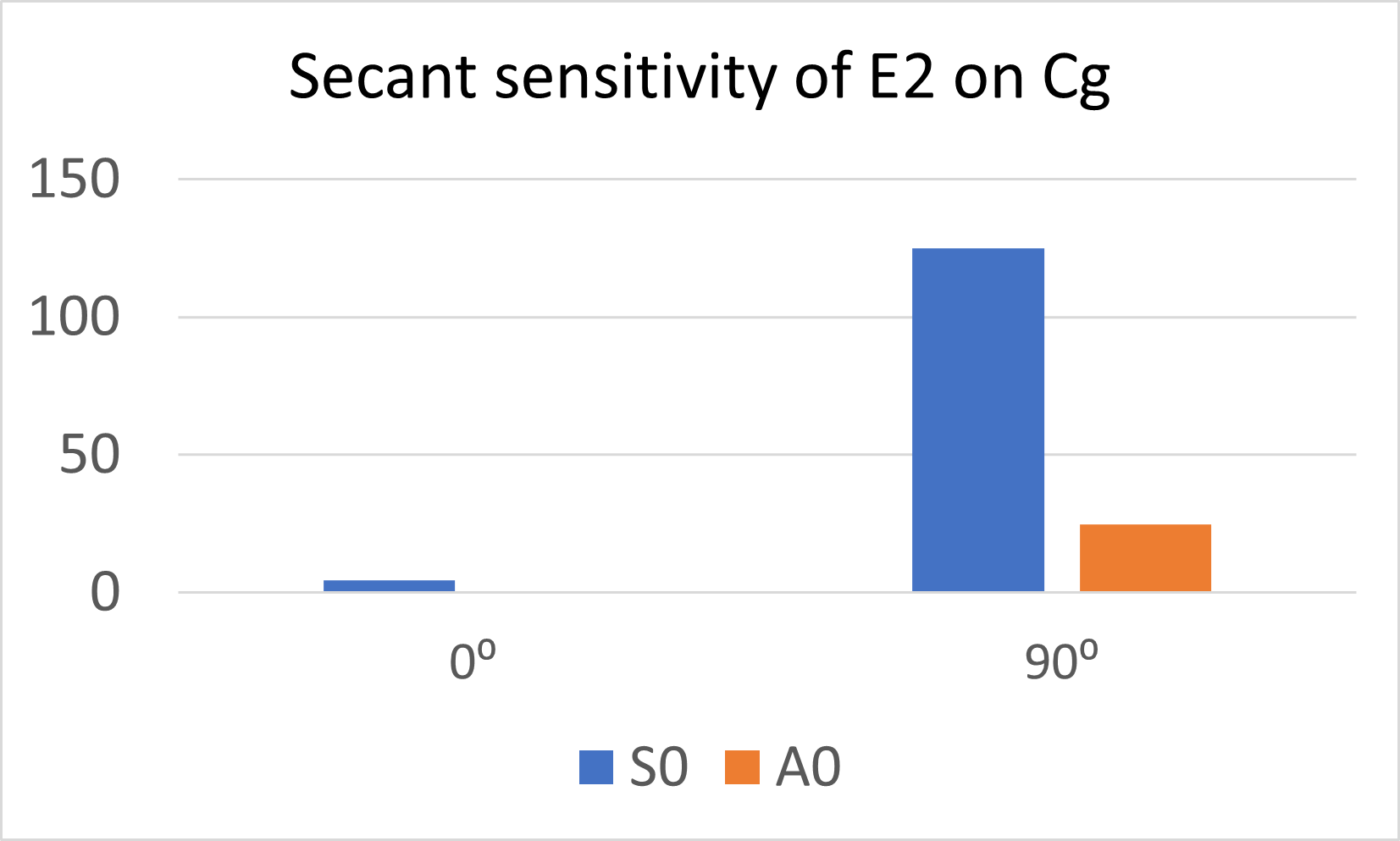}
	%\subcaption{$v_{12}$ effect on SH0}
	\vspace{1ex}
\end{minipage}
\centering
\begin{minipage}[b]{0.32\textwidth}
	\centering
	\includegraphics[width=1.0\textwidth]{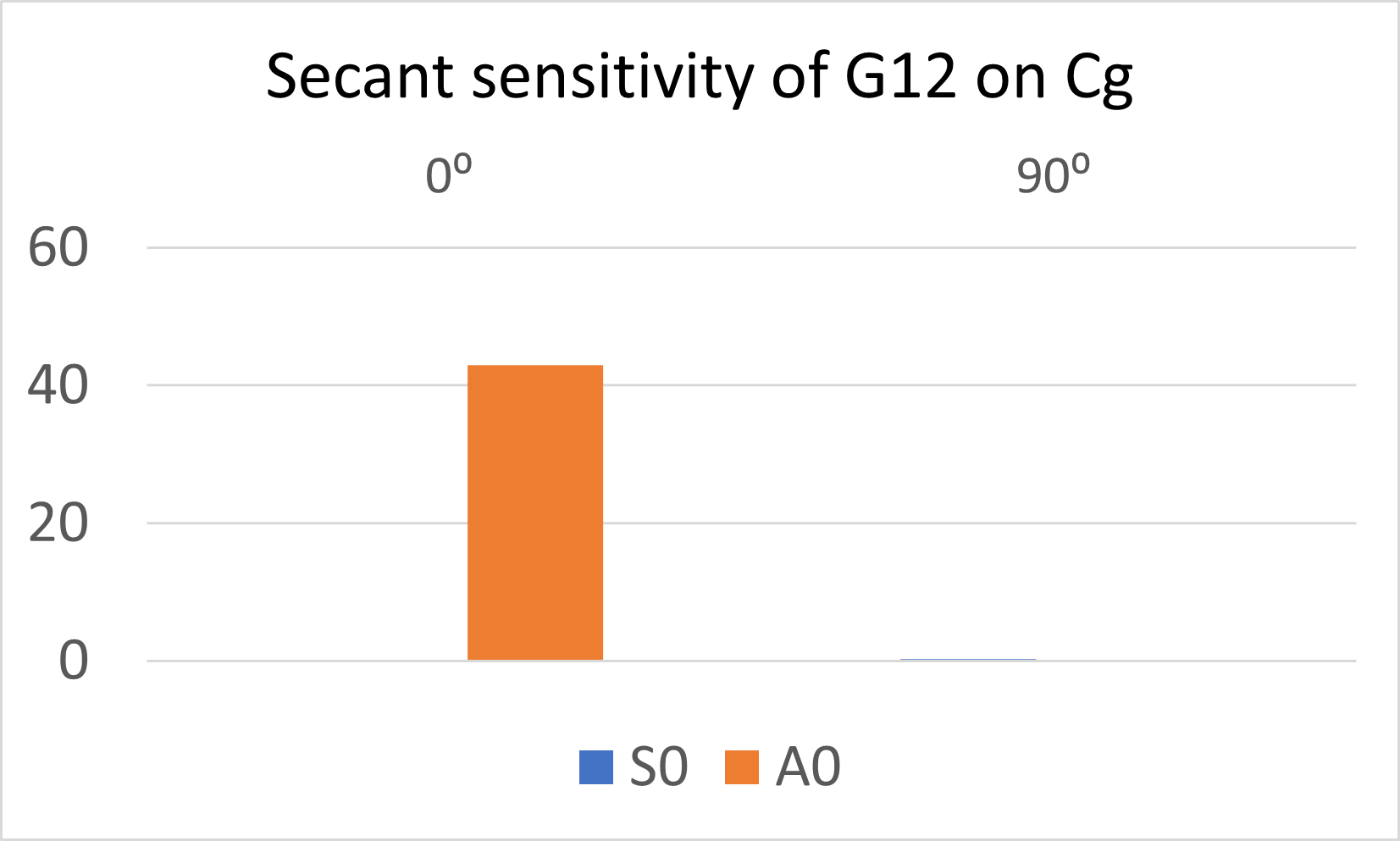}
	%\subcaption{$v_{23}$ effect on $S_0$}
	%\vspace{4ex}
\end{minipage}
\begin{minipage}[b]{0.32\textwidth}
	\centering
	\includegraphics[width=1.0\textwidth]{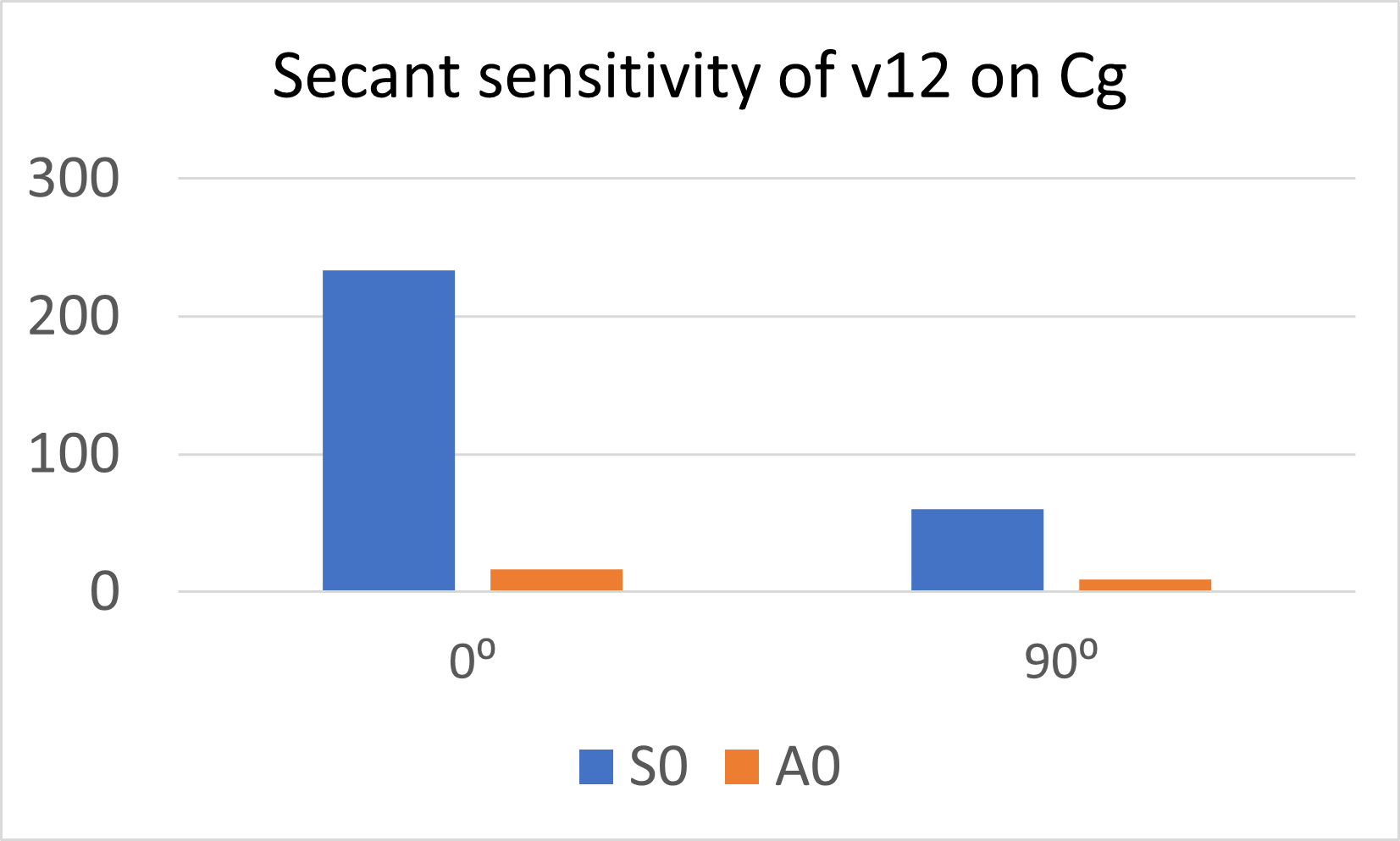}
	%\subcaption{$v_{23}$ effect on A0}
	%\vspace{4ex}
\end{minipage}
\begin{minipage}[b]{0.32\textwidth}
	\centering
	\includegraphics[width=1.0\textwidth]{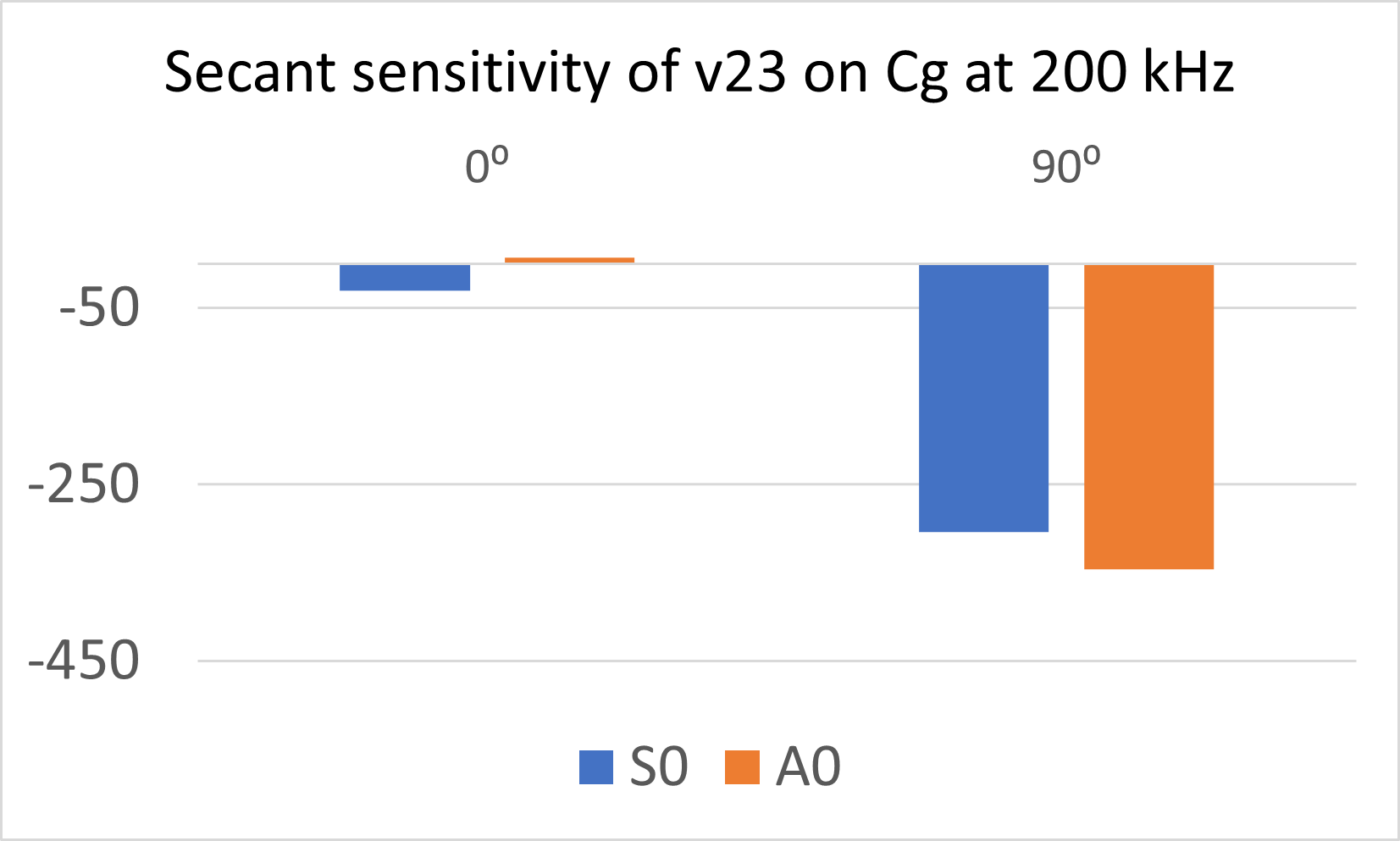}
	%\subcaption{$v_{23}$ effect on SH0}
	%\vspace{4ex}
\end{minipage}
\caption{\small Secant sensitivity of $\rho$, $E_1$, $E_2$, $G_{12}$, $\nu_{12}$, $\nu_{23}$ on group velocities of $S_0$ and $A_0$ modes at 0$^\circ$ and 90$^\circ$ propagation angles.}
\label{fig:secant}
\end{figure}

It is seen that different properties have various kinds of effects on polar group velocities. Each property is influencing at least one of the modes at one or more than one frequency. The effects are more visualized for $\rho$, $E_1$, $E_2$ and $G_{12}$ than $\nu_{12}$ and $\nu_{23}$. All the plots of Fig.~\ref{fig:sensivitiy} are presented for 20 kHz except for $\nu_{23}$ which is plotted at 200 kHz since the changes are very minimal for $\nu_{23}$ at lower frequencies. The benefit of including multiple modes and frequencies in this study is evident from sensitivity analysis. In order to give a quantitative measure of sensitivity, secant sensitivity is used to calculate the effect of different parameters on the group velocities using Eq.~(\ref{eq:secant}).

\begin{equation}\label{eq:secant}
\delta = \frac{C_e - C_s}{P_e-P_s}
\end{equation}
where $C_e$ is the end value of the group velocity in m/s for the lamina property $P_e$ and $C_s$ is the start value of the group velocity for $P_s$. For density, $P_s$ = 1304 kg/$\mathrm{m}^3$ and $P_e$ = 1760 kg/$\mathrm{m}^3$ whereas $C_s$ and $C_e$ are the corresponding group velocities. The secant sensitivity for the group velocities of both modes at the propagation angles of 0$^\circ$ and 90$^\circ$ are plotted in Fig.~\ref{fig:secant}.

Negative values of secant sensitivity in the figure represent the decreasing trend of the parameter on the velocity whereas positive values correspond to an increasing trend. Here, secant sensitivity is used to compare velocities of both modes along with two different propagation angles for each material property. Due to the different S.I. units of the material properties, any comparison across different material properties is restricted. It is observed from Figs.~\ref{fig:sensivitiy} \& \ref{fig:secant} that $\rho$, $E_1$ and $\nu_{12}$ have a higher effect on the $S_0$ mode than on the $A_0$ mode. These properties have a greater influence at 0$^\circ$ than at 90$^\circ$ propagation angle. This relationship is reversed for $\nu_{23}$. $E_2$ has a dominating effect for the $S_0$ mode and 90$^\circ$ propagation angle. $G_{12}$ shows a higher effect at 0$^\circ$ than at 90$^\circ$.

%------------------------ Begin NEW SECTION
\section{Training strategy for the networks}\label{sec:train}
\subsection{Data collection} \label{ssec:data}
Two material sets (Matsets) are generated for this study. Matset-1 is a set which contains the materials mentioned in Table - \ref{tab:ComMat}. For Matset-2, the six material properties are varied in a range ($\rho$ = [1304 1760] kg/$\mathrm{m}^3$, $E_1$ = [115 184] GPa, $E_2$ = [6 14] GPa, $G_{12}$ = [3 9] GPa, $\nu_{12}$ = [0.2 0.52], $\nu_{23}$ = [0.23 0.59]). A vector of six elements which corresponds to the six material properties is randomly selected from a uniform distribution with the abovementioned bounds. Each material (Matset-1 \& 2) is fed into the forward model (See Fig.~\ref{fig:forward}), which provides polar group velocities at different excitation frequencies ranging from 20 kHz to 200 kHz in increments of 20 kHz. Both the fundamental lamb modes are dispersive in this frequency range. This range is considered suitable for Lamb-wave propagation-based experiments because the selected range with the given thickness eliminates higher lamb wave modes which may complicate the study \cite{mitra2016guided}. This process is performed for three different ply-layup sequences, i.e., unidirectional, cross-ply, and quasi-isotropic with layup symmetry and 16 layers with 2 mm thickness. Further, the polar group velocities are transformed into binary images (black \& white images) called polar representations. Two datasets corresponding to two different material sets are collected.

For the first inverse problem, the polar representations are classified into three different layup sequence classes. Here, dataset-1 is used, which contains 100 samples (10 materials with 10 different frequencies) per class for each Lamb mode (or branch). The overall dataset consists of 300 samples. For the second inverse problem, the polar representations are utilized as input to predict six material properties using a regression model. For this, a larger dataset (dataset-2) is generated, which includes 10,000 samples (1000 materials with 10 different frequencies) per branch. Different sized datasets are used for both problems because identifying material properties is a more complex inverse problem than the classification of the ply layup sequence. In order to demonstrate the inversion capability of our material property identification model, the training is performed with a single layup sequence type.

\subsection{Featurization of polar representations}\label{ssec:FeatureEngg}
\begin{figure}
%\captionsetup{justification=centering}
\centering
\begin{minipage}[b]{1.0\textwidth}
	\centering
	\includegraphics[width=1.0\textwidth]{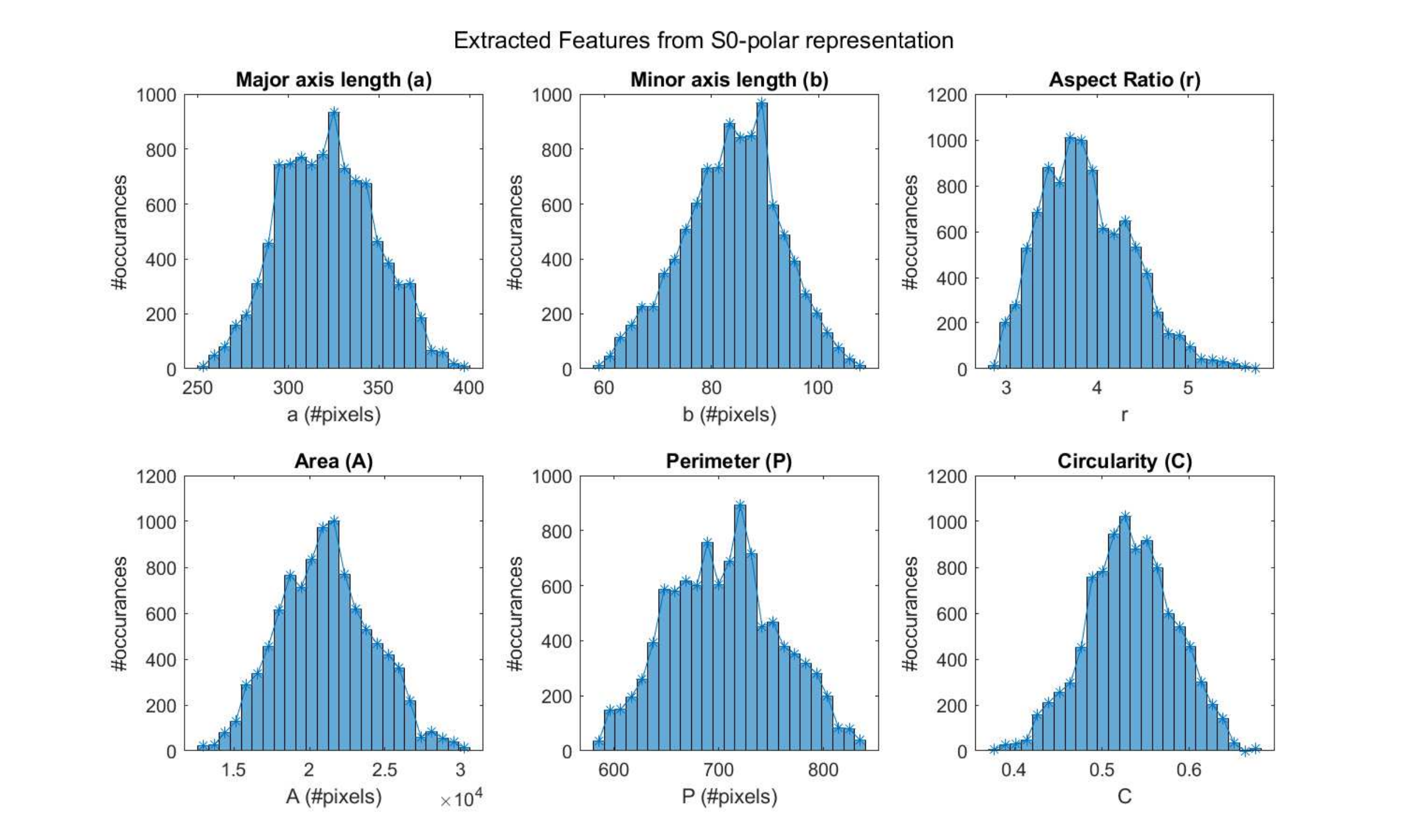}
	\vspace{1ex}
\end{minipage}
\begin{minipage}[b]{1.0\textwidth}
	\centering
	\includegraphics[width=1.0\textwidth]{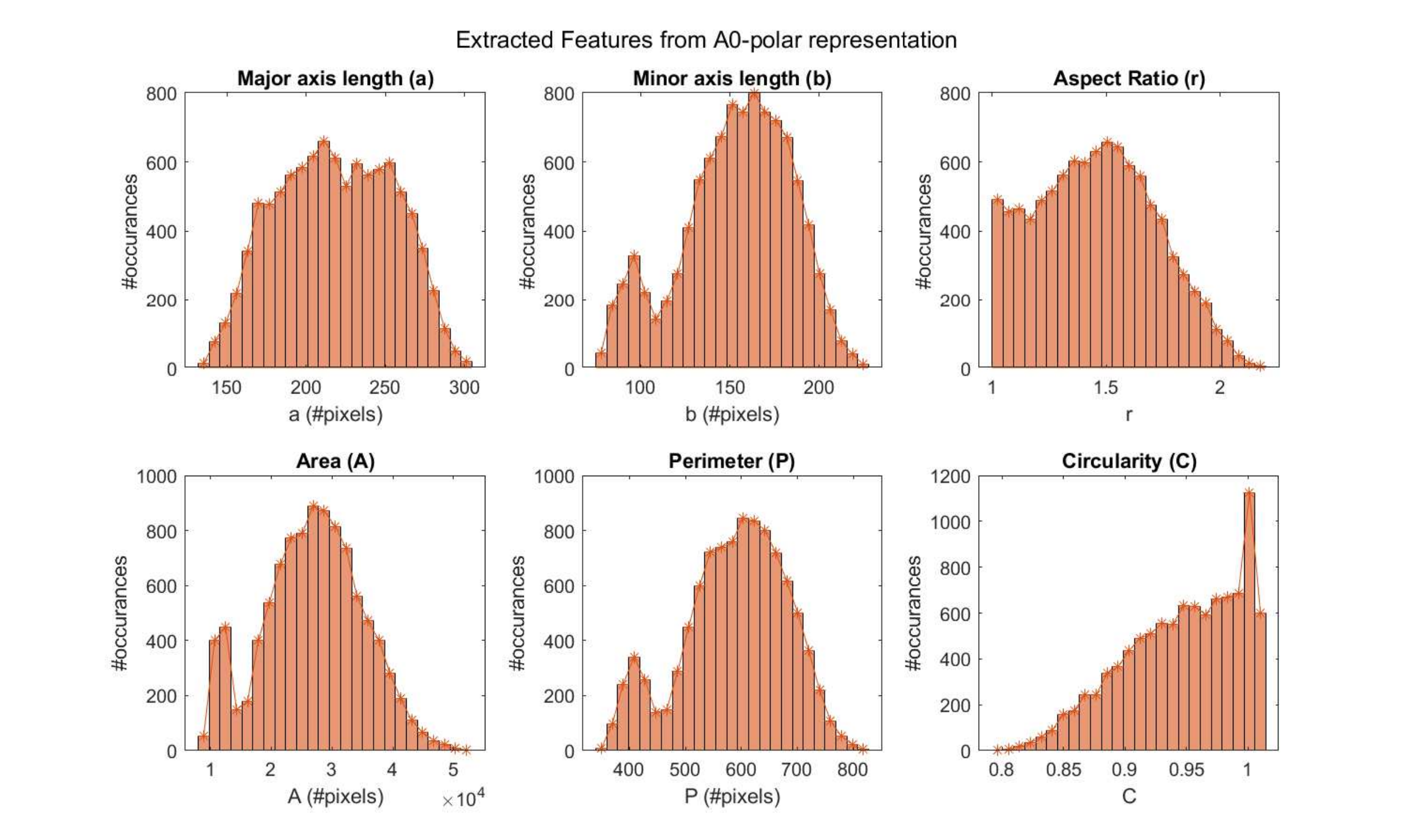}
	%\vspace{4ex}
\end{minipage}
\caption{\small Distribution of features extracted from the polar representations of $S_0$ and $A_0$ modes across the dataset.}
\label{fig:FeatureEngg}
\end{figure}

Featurization or feature-engineering is recommended for larger datasets, even if the workflow involves automatic feature extraction from the representations via deep learning. It helps in understanding the dataset and behavior of signals. It also encourages the implementation of machine learning models for comparisons. Features such as major-axis length (a), minor-axis length (b), aspect-ratio (a/b), area (A), perimeter (P), and circularity (C=4$\pi$A/$P^2$) are extracted from the polar representations of both the modes present in the dataset-2. These quantities are presented in the form of the number of pixels in 600$\times$600 resolution binary images. The distribution of the features in the dataset is presented in the form of histograms for both A0 and S0 modes in Fig.~\ref{fig:FeatureEngg}. For this, A rectangular bin of width 25 is selected. In the figure, the ordinate shows the number of occurrences of the bin in the dataset for each feature. The summation of all occurrences is bounded by the size of the dataset.

From the figure, various observations can be made about the dataset-2. The difference between major and minor axis (a-b) for $S_0$-polar representations is more than for $A_0$-polar representations, which directly reflects on higher circularity and lower aspect ratio of $A_0$-polar representations. One important point to note down here is that it is extraneous to compare feature-by-feature across both types of representations (for instance, comparing a, b, A, P across both the representations). This is because the $A_0$ polar plots are scaled-up for better training purpose due to a higher difference between group velocities of both modes.

This featurized dataset is used to train machine learning algorithms in Sec.~\ref{ssec:compare}. The results of these algorithms are compared against our proposed approach.

\subsection{Cost function, optimization scheme and metrics} \label{ssec:cost}

The selection of a proper cost function is important in neural networks based learning scheme. For this, a gradient descent-based optimization algorithm is utilized to drive the cost function towards the optima. Cross-entropy loss function is best suited where the outputs are discrete (ply layup sequence type) and mean squared error is utilized where the outputs are continuous values (material properties) \cite{hampshire1990novel,goodfellow2016deep}. In this work, a categorical cross-entropy cost function given by Eqs.~(\ref{eq:cce}) is selected for the classification problem. It is similar to the binary cross-entropy loss but with a sigmoid activation in the last layer is replaced by a softmax activation function. A mean-square error is used for the regression problem, which is mathematically represented by Eq.~(\ref{ref:mse}). 

%---equation
\begin{equation}\label{eq:cce}
J_c(W,b) = \frac{1}{m} \sum_{i=1}^{m} (y^i\log\hat{y}^i + (1-y^i)\log(1-\hat{y}^i))
\end{equation}

%---equation
\begin{equation}\label{ref:mse}
J_r(W,b) = \frac{1}{m} \sum_{i=1}^{m} (y^i-\hat{y}^i)^2
\end{equation}

where $m$ is the total number of training examples, $W$ are the weights and $b$ are the biases of the network, $y$ and $\hat{y}$ are the true and predicted output, respectively.

The learning parameters (\textit{W,b}) are updated through a batch gradient descent algorithm-based optimization scheme using a back-propagation technique as shown in Eqs.~(\ref{eq:W}) and (\ref{eq:b}). An Adam optimizer is used to enable momentum and adaptive learning rate into the training process \cite{kingma2014adam}. 

%---equation
\begin{equation}\label{eq:W}
W_{j+1} = W_{j}-\alpha\frac{\partial J(W,b)}{\partial W}\bigg|_{W=W_j}
\end{equation}
%---equation
\begin{equation}\label{eq:b}
b_{j+1} = b_{j}-\alpha\frac{\partial J(W,b)}{\partial b}\bigg|_{b=b_j}
\end{equation}
where $j$ is the iteration, $\alpha$ is the learning rate which decides the step-size during the course of iterations.

Metrics are used to assess the training and validation performance of the networks. For classification, accuracy is used as a metric as shown in Eq.~(\ref{eq:acc}).
%-------
\begin{equation}\label{eq:acc}
\text{A} = \frac{\text{TP + TN}}{\text{TP + TN + FP + FN}}
\end{equation}
where TP = True Positives, TN = True Negatives, FP = False Positives, and FN = False Negatives.

For the regression problem, Mean Absolute Percentage Error (MAPE) and coefficient of determination ($R^2$ value) are selected as the metrics as represented mathematically in Eqs.~(\ref{eq:mape}) \& (\ref{eq:r2}). MAPE gives the overall prediction error in percentage whereas the $R^2$ value provides a `goodness of fit'.

%---equation
\begin{equation}\label{eq:mape}
M = \frac{100}{m} \sum_{i=1}^{m} \bigg|\frac{y^i-\hat{y}^i}{y^i}\bigg|
\end{equation}

%---equation
\begin{equation}\label{eq:r2}
R^2 = 1 - \frac{\sum_{i=1}^{m} (y^i-\hat{y}^i)^2}{\sum_{i=1}^{m} (y^i-\bar{y})^2}
\end{equation} 

where, $m$ is the total number of training examples, $y$ and $\hat{y}$ is the true and predicted output, and $\bar{y}$ is the mean output.

The networks are trained on a GPU of Nvidia RTX-2070 (CUDA cores = 2304 and Tensor Cores = 288) with 8 GB of VRAM and a CPU configuration of i7-9700KF (8 cores) with 32 GB RAM. The code files are developed in a TensorFlow and Keras environment using Python programming language and the code package is available on Github.

%---------------------- Begin NEW SECTION
\section{Identification of layup sequence type}\label{sec:findply}
The inverse problem of the ply layup sequence is designed as a classification problem where CNN with dual-branch feature fusion (dual-branch CNN) is used to classify the layup sequence types into three classes, i.e., unidirectional, cross-ply, and quasi-isotropic layup. Dataset-1 contains 100 samples per class for each branch (overall dataset size is 300 per branch). The dataset is randomly split into training and validation sets based on a 9:1 split. Training examples used to train the networks, whereas other examples are the unseen examples used only in the validation and testing phase. Each training example is associated with a label, i.e., 0 for unidirectional laminate, 1 for cross-ply laminate, and 2 for quasi-isotropic laminate. These labels are converted into one hot-encoded representations, i.e., a vector of 1$\times$3 for each training example.

The networks are trained with a learning rate of 0.001 and a batch size of 4. These hyperparameters are selected based on the smoothness of the loss curve and the generalization in the testing phase. The architecture of the trained network for the classification model is shown in Table-\ref{tab:classificationmodel}. Two parallel branches of CNN are used with a similar architecture. Each branch consists of three convolutional layers with 16, 32, and 64 filters, respectively, in each layer. Rectified Linear Unit (ReLU) is used as an activation function in each layer. A square kernel window with size 3$\times$3 is used in all convolutional layers. Each convolutional layer is followed by a batch normalization unit and a max-pooling layer with size 2$\times$2 to downsample the output of the convolution layer. The features in the respective parallel branch are flattened out into a single vector. These features are fused together, which increases the number of the features by two times. An FCN is used with 16 neurons and ReLU activation followed by an output layer of 3 neurons with a softmax activation. The total number of training parameters for the classification model is 0.57 million.

\begin{table}
\centering
\captionsetup{justification=centering}
\setlength{\belowcaptionskip}{0pt}
\caption{\small Architecture of the dual-branch CNN based classification model for identification of ply-layup sequence type.}
\addtolength{\tabcolsep}{-1pt} 
\begin{tabular}{l|c|c}
	\hline
	Layer & Output Shape & Parameters\\
	\hline
	\textbf{CNN-1\&2:} &&\\
	Input Layer & (128,128,1) & 0 \\
	Conv2D (ReLU, 3x3, 16) $\rightarrow$ BatchNorm $\rightarrow$ MaxPool (2x2) & (64,64,16) & 224 \\
	Conv2D (ReLU, 3x3, 32) $\rightarrow$ BatchNorm $\rightarrow$ MaxPool (2x2) & (32,32,32) & 4768 \\
	Conv2D (ReLU, 3x3, 64) $\rightarrow$ BatchNorm $\rightarrow$ MaxPool (2x2) & (16,16,64) & 18752 \\
	Flatten() & (16384) &  0 \\
	\hline
	\textbf{Feature Fusion:} & (32768) & 0\\
	\hline
	\textbf{FCN:} &&\\
	Dense(ReLU, 16) & (16)  & 524,304 \\
	Dense(Softmax, 3) & (3) & 51 \\
	\hline
	Trainable parameters &-& 571,395 \\
	Non-trainable parameters &-& 448 \\
	Total parameters &-& 571,843 \\
	\hline
\end{tabular}
\label{tab:classificationmodel}
\end{table}

\begin{figure}
%\captionsetup{justification=centering}
\centering
\begin{minipage}[b]{1.0\textwidth}
	\centering
	\includegraphics[width=0.75\textwidth]{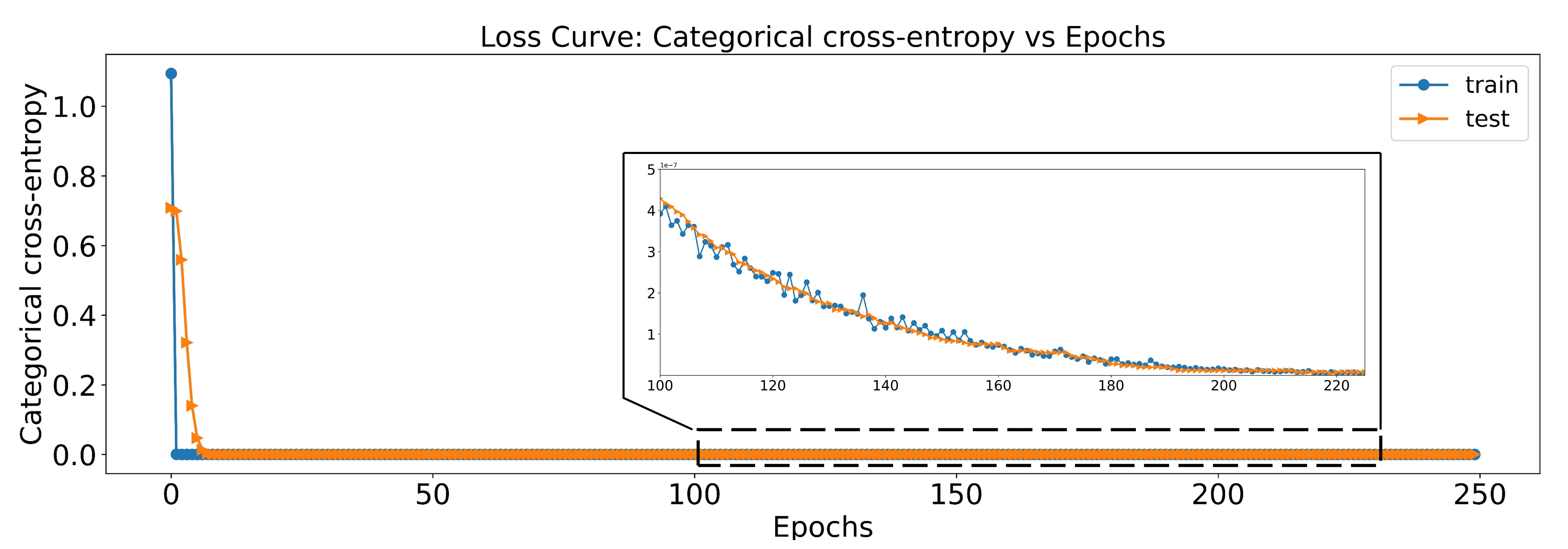}
	%\subcaption{S0 mode}
	%\vspace{4ex}
\end{minipage}
\begin{minipage}[b]{1.0\textwidth}
	\centering
	\includegraphics[width=0.875\textwidth]{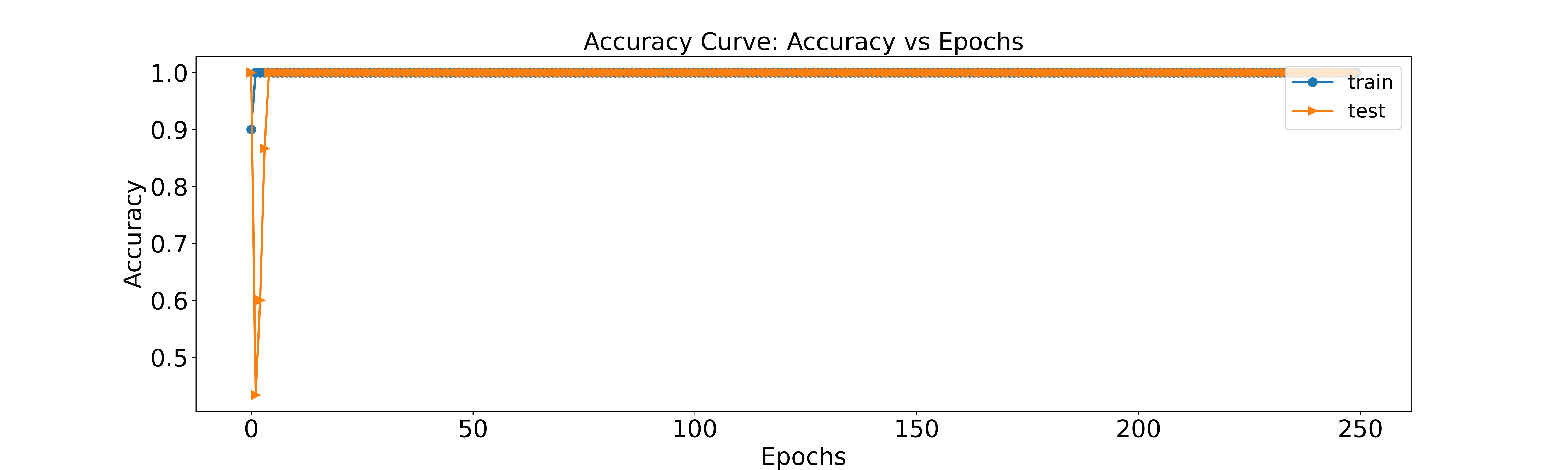}
	%\subcaption{A0 mode}
	%\vspace{4ex}
\end{minipage}
\caption{\small (a) Loss curve and (b) accuracy curve for dual-branch CNN model for classification of polar representations in ply layup types (unidirectional, cross and quasi-isotropic).}
\label{fig:results_class}
\end{figure}

For the classification model, categorical cross-entropy loss given by Eq.~(\ref{eq:cce}) is used as a cost function with the Adam optimization scheme. The network is trained for 250 epochs. An epoch scheduler is used as a callback function on loss value and accuracy. The training time per epoch is 1 second with a total of 250 epochs. The training process is repeated ten times while randomly shuffling the dataset to ensure the repeatability of the results. The testing time for each unseen example is less than one millisecond. The loss and accuracy curves are presented in Fig.~\ref{fig:results_class}. It is seen that the network is able to obtain a training loss of 8e-9 with an accuracy of 1.0 and a validation loss of 9e-9 with an accuracy of 1.0.

%\begin{figure}[t]
%\centering
%\includegraphics[width=.25\textwidth]{confusionmat.png}
%\caption{Confusion matrix}
%\label{fig:forward}	
%\end{figure}

%-------------------- NEW SECTION
\section{Identification of material properties}\label{sec:findproperties}
Dataset-2 is used for the identification of material properties containing 10,000 samples per branch in the form of polar representations. The dataset is randomly split into training and test sets, the training set consists of 8519 examples and the test set consists of 1481 examples. Each training example is associated with six material properties ($\rho$, $E_1$, $E_2$, $G_{12}$, $\nu_{12}$, $\nu_{23}$) as labels. A dual-branch CNN based regression network is used for the identification of six material properties.

A learning rate of 1e-5 and a batch size of 16 is used to train the network. The architecture of the network for the regression model is shown in Table-\ref{tab:regressmodel}. Here also, Two parallel branches of CNN are used with a similar architecture. Each branch has four convolutional layers with 16, 32, 64, and 128 filters with a ReLU activation function. Kernel-size, Batch-Norm, Max-Pooling are similar to the classification network. The features are fused, followed by a single layer of FCN having 256 neurons with ReLU activation. The output layer with six neurons and a linear activation is used. A dropout layer with a drop rate of 13\% is used to incorporate regularization and reduce the overfitting of the model. The total number of training parameters for the regression model is 4.39 million.

\begin{table}[t!]
\centering
\captionsetup{justification=centering}
\setlength{\belowcaptionskip}{0pt}
\caption{\small Architecture of the dual-branch CNN based regression model for identification of material properties.}
\addtolength{\tabcolsep}{-1pt} 
\begin{tabular}{l|c|c}
	\hline
	Layer & Output Shape & Parameters\\
	\hline
	\textbf{CNN-1\&2:} &&\\
	Input Layer & (128,128,1) & 0 \\
	Conv2D (ReLU, 3x3, 16) $\rightarrow$ BatchNorm $\rightarrow$ MaxPool (2x2) & (64,64,16) & 224 \\
	Conv2D (ReLU, 3x3, 32) $\rightarrow$ BatchNorm $\rightarrow$ MaxPool (2x2) & (32,32,32) & 4768 \\
	Conv2D (ReLU, 3x3, 64) $\rightarrow$ BatchNorm $\rightarrow$ MaxPool (2x2) & (16,16,64) & 18752 \\
	Conv2D (ReLU, 3x3, 128) $\rightarrow$ BatchNorm $\rightarrow$ MaxPool (2x2) & (8,8,64) & 74368 \\
	Flatten() & (10368) &  0 \\
	\hline
	\textbf{Feature Fusion:} & (20736) & 0\\
	\hline
	\textbf{FCN:} &&\\
	Dense(ReLU, 256) $\rightarrow$ Dropout(13\%) & (256)  & 4,194,560 \\
	Dense(Linear, 6) & (6) & 1542 \\
	\hline
	Trainable parameters &-& 4,391,366 \\
	Non-trainable parameters &-& 960 \\
	Total parameters &-& 4,392,326 \\
	\hline
\end{tabular}
\label{tab:regressmodel}
\end{table}

\begin{figure}
%\captionsetup{justification=centering}
\centering
\begin{minipage}[b]{1.0\textwidth}
	\centering
	\includegraphics[width=0.7\textwidth]{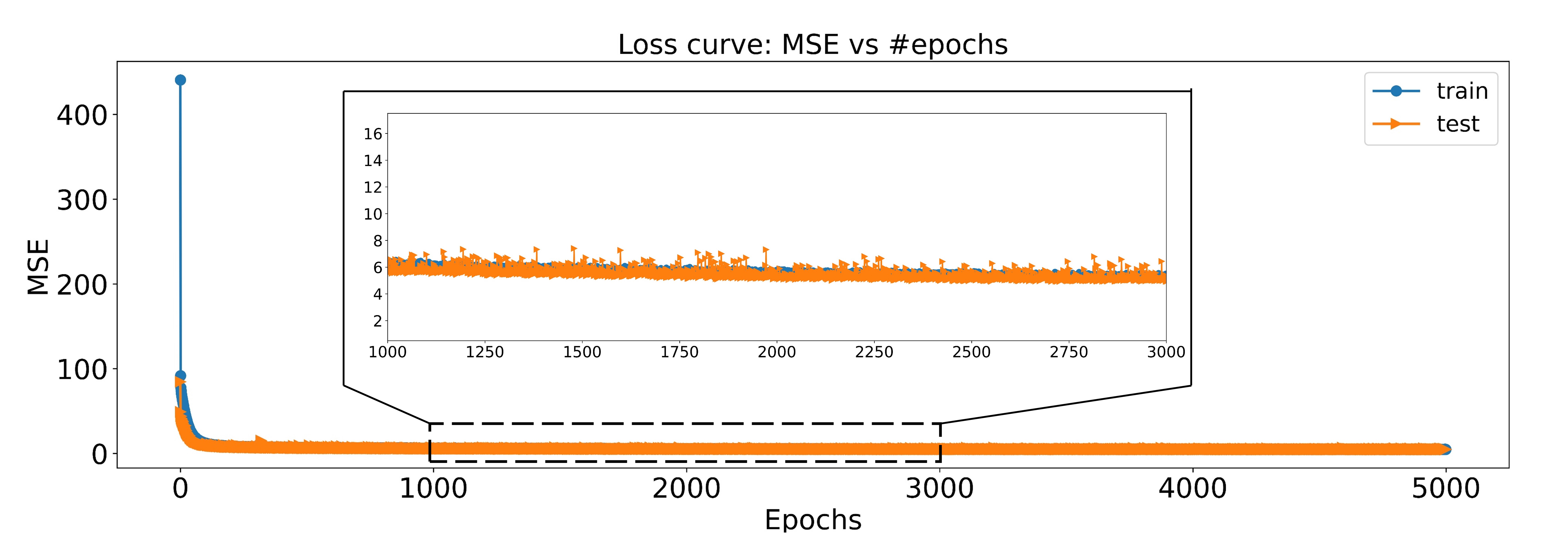}
	\subcaption{Loss curve: MSE vs Epochs}
	\vspace{1ex}
\end{minipage}
\begin{minipage}[b]{1.0\textwidth}
	\centering
	\includegraphics[width=0.8\textwidth]{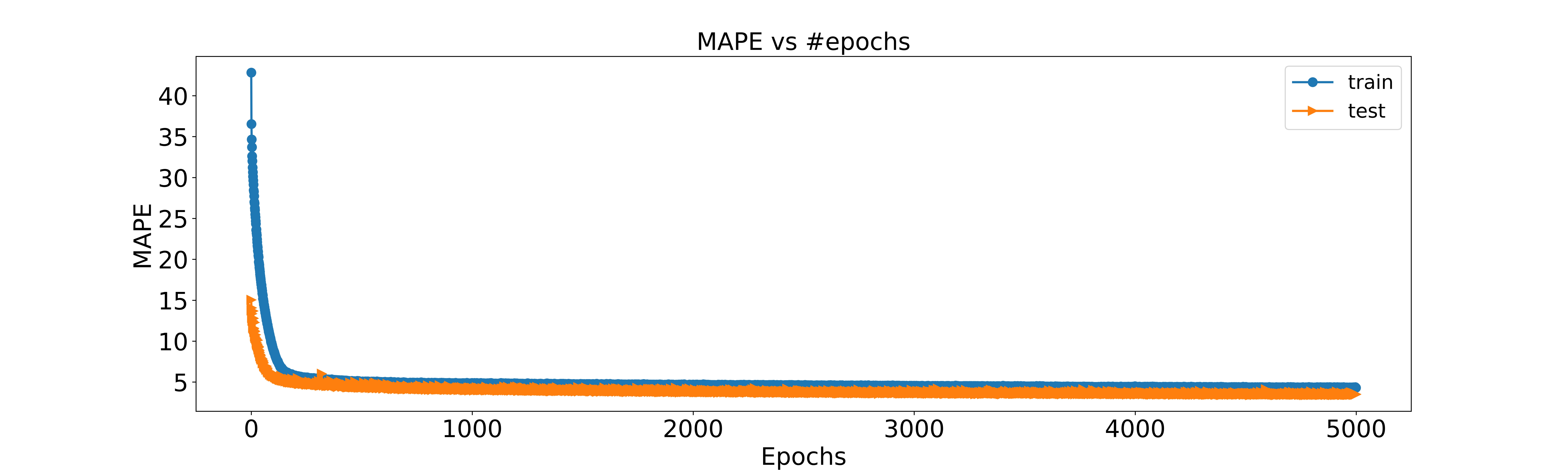}
	\subcaption{Mean absolute percentage error vs Epochs}
	\vspace{1ex}
\end{minipage}
\begin{minipage}[b]{1.0\textwidth}
	\centering
	\includegraphics[width=0.75\textwidth]{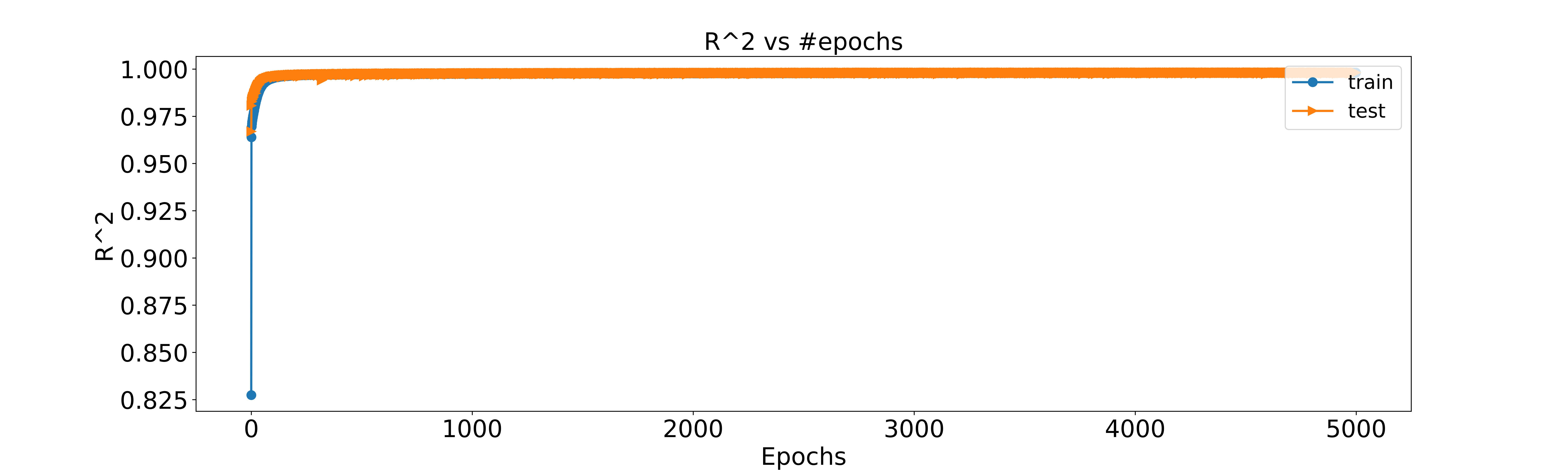}
	\subcaption{Coefficient of determination ($R^2$) vs Epochs}
	\vspace{1ex}
\end{minipage}
\caption{\small MSE Loss, MAPE, and $R^2$ curves for dual-branch CNN based regression model for identification of material properties}
\label{fig:results_regress}
\end{figure}

For the regression model, mean squared error presented in Eq.~(\ref{eq:cce}) is used as a cost function with the Adam optimization scheme. The network is trained for 5000 epochs with an epoch scheduler on loss and MAPE. The training time is 16 seconds per epoch with a total of 5000 epochs. Similar to the training of the classification model, the training process is repeated ten times to ensure the repeatability. MSE loss, MAPE, and $R^2$ curves are shown in Fig.~\ref{fig:results_regress}. An MSE of 5.0 and 4.65, MAPE of 4.0 and 3.5, $R^2$ of 0.998 is achieved in training and cross-validation, respectively.

An important point to highlight here is related to the selection of network architecture for both classification and regression problems in hand. We have followed a trial and error-based method where the number of neurons and depth of the network is increased gradually along with other hyperparameters satisfying minimum loss values, reduced overfitting and underfitting, smoother loss curves, reduced computational time and better generalization on the test set.

%---------------------- NEW SECTION
\section{Testing results and comparisons}\label{sec:results}
\subsection{Testing the networks}\label{ssec:test}
A confusion matrix analysis is performed for the classification model and seen that all 30 test examples (uni=10, cross=11, and quasi=9) are classified accurately into their respective classes. The trained model is also tested on dataset-2, containing polar representations of a unidirectional layup. As mentioned earlier, the polar representations of dataset-2 come from the material properties randomly sampled from a uniform distribution. The model has accurately classified all the 10,000 representations present in dataset-2. This verifies the success of the approach for the classification of the ply-layup sequence type. 

The trained regression model is utilized to perform prediction on unseen examples (1481 examples) present in the test set. Along with this, the trained network is tested on dataset-1, which includes commercial composite materials and their polar representations. The results in the form of MAPE between the true value of the material properties and the predicted values from the network are tabulated in Table-\ref{tab:DL}.

\begin{table}[h!]
\centering
\captionsetup{justification=centering}
\setlength{\belowcaptionskip}{0pt}
\caption{\small Test results: Mean absolute percentage error (MAPE) on Dataset-1 and 2}
\addtolength{\tabcolsep}{-1pt} 
\begin{tabular}{c|c c c c c c}
	\hline
	Dataset & $\rho$ & $E_1$ & $E_2$ & $G_{12}$ & $\nu_{12}$ & $\nu_{23}$\\
	\cline{1-7}
	Dataset-2  & 2.0 & 1.6 & 2.9 & 3.6 & 5.4 & 5.4 \\
	Dataset-1  & 1.6 & 1.3 & 2.3 & 3.4 & 4.7 & 5.1 \\
	\hline
\end{tabular}
\label{tab:DL}
\end{table}

It is seen that the results from both networks are promising. The classification-based model trained on dataset-1 is able to classify unseen test samples of polar representations into one of the three layup sequences with high accuracy. The model has proved its prediction ability on polar representations of dataset-1, which is created based on random sampling. The regression-based model trained on dataset-2 has performed very well in prediction on unseen samples with a maximum of 5.4\% MAPE. The trained model has performed outstandingly on commercially available CFRP composite materials (dataset-1). The remarkable testing results of both the models on different datasets prove the generalization ability of the trained models. 

The prediction time for the classification and regression model is in the order of milliseconds per sample. It ensures the usage of such models for in-situ monitoring of degradation of material properties in extreme environments, generally faced in applications like aerospace. This advantage in prediction time is helpful in rapid non-destructive material property measurements in mass-scale production systems. Fast predictions and real-time deployment is one of the major drawbacks of the heuristics-based inversion schemes presented in the literature.

\subsection{Comparisons \& Discussions}\label{ssec:compare}
A featurized dataset is generated with features such as major axis length (a), minor axis length (b), aspect ratio (r), area (A), perimeter (P), and circularity (C) in terms of the number of pixels for the dataset-2 (Refer Sec.~\ref{ssec:FeatureEngg} with Fig.~\ref{fig:FeatureEngg}).Supervised machine learning (ML) based algorithms like SVM \cite{pan2018time}, Linear-Lasso-Ridge regression \cite{shahidi2015structural}, Random Forest \cite{zhou2014structure}, and ANN \cite{zhao1998structural} applied on the featurized dataset. Apart from presenting a different philosophy (ML based) to solve the inverse problem of material characterization, this exercise is also performed to compare the ML models against our deep learning (DL) technique. Both ML and DL methods are different in terms of the feature extraction process. Generally, ML uses domain expertise to featurize the data, whereas DL performs automatic feature extraction on the representations. However, the features extracted in DL may be less explainable than ML, but with DL, the extracted features are complex and more representative at different levels of abstraction. Such an automatic feature process plays a dominant role in solving complex engineering problems \cite{rautela2021ultrasonic}. 

During the training process of ML models, the dataset is split randomly into two parts, with nearly 8519 training samples and 1481 test samples. Three different versions of SVM are implemented with kernels like Radial Bias Function (SVM-RBF), Linear, Polynomial of degree 3 (SVM-P3). The values of gamma (radius of the area of influence of the support vectors) are selected as 0.1 for RBF and are chosen automatically for the other two. The regularization or penalty parameter, C, is chosen to be 10. For the Ridge and Lasso regression models, alpha (regularization parameter) is set to 20 and 2, respectively. 30 trees are used in Random Forest. For ANN, 2048 neurons with ReLU and linear activation are selected. The learning rate of 0.001 and batch-size of 64 is selected and the network is trained for 1000 epochs. A thorough description about these ML models are explained in Ref.~\cite{scikit-learn}. The code files for ML models are developed in the Python programming language, and the code package is open-sourced on Github. The prediction results of the trained algorithms on the test set is presented in Table-\ref{tab:ML}. 

\begin{table}[h!]
	\centering
	\captionsetup{justification=centering}
	\setlength{\belowcaptionskip}{0pt}
	\caption{\small MAPE obtained from ML algorithms applied on featurized dataset}
	\addtolength{\tabcolsep}{-1pt} 
	\begin{tabular}{c|c c c c c c c c}
		\hline
		Property  & SVM(RB) & SVM(L) & SVM(P3) & Linear & Ridge & Lasso & RF & ANN\\
		\hline
		$\rho$     & 5.5  & 5.6  & 5.5  & 5.6  & 5.6  & 7.5  & 2.8 & 8.3 \\
		$E_1$      & 5.6  & 5.6  & 5.6  & 5.6  & 5.7  & 9.0  & 2.6 & 5.7 \\
		$E_2$      & 5.9  & 6.8  & 5.9  & 6.8  & 7.1  & 22.9 & 3.3 & 11.4 \\
		$G_{12}$   & 5.8  & 6.8  & 5.8  & 6.9  & 8.5  & 28.2 & 3.9 & 11.0 \\
		$\nu_{12}$ & 22.0 & 22.0 & 21.8 & 22.3 & 22.9 & 23.2 & 7.6 & 22.8 \\
		$\nu_{23}$ & 23.2 & 23.0 & 23.0 & 22.9 & 23.3 & 23.5 & 8.7 & 23.8 \\
		\hline
	\end{tabular}
	\label{tab:ML}
\end{table}

The test results in the form of MAPE can be used to compare our proposed technique (Table-\ref{tab:DL}) against the ML models (Table-\ref{tab:ML}). It is seen that the dual-branch CNN model achieve much better results than the ML models on the test set of dataset-2. Out of eight algorithms, the results from the Random Forest approach are closer to the DL results for dataset-2. However, the prediction on dataset-1 is falling behind by a larger margin.

In this paper, we have solved the inverse problem of composite material characterization where the datasets are the outputs of a forward computational model. The networks used to solve the inverse problems are based on polar group velocity representations (See Figure-\ref{fig:inverse}) as inputs. However, the experimental process includes time-series signals coming from a circular array of piezoelectric sensors mounted on a composite structure. The group velocity calculations from experimental signals is a well-studied problem in the literature with established ways of solving it \cite{wang2007group,su2009identification,mitra2016guided,malik2021direct}. The group velocity can be calculated corresponding to each sensor’s signal manually or automatically by analyzing the waveform even in the presence of noise. Polar group velocity representations can be generated using group velocities as a function of propagation angle. The representations can be used to train a network from the beginning if sufficient experimental observations are available. In the setting of our current approach, the network trained on representations generated from a computational model can be utilized to predict the material properties and ply-layup type corresponding to representations coming either from the computational model or from experimental observations.

% ------------------------ NEW SECTION
\section{Conclusions}\label{sec:conclusion}
In this paper, a deep supervised learning approach is implemented to solve the inverse problem of the composite material characterization. The stiffness matrix method along with group velocity calculation, is used to solve the forward problem in which polar group velocity representations are obtained as outputs while material properties and ply-layup sequences are the inputs. An in-depth sensitivity analysis is presented on the forward model to understand the contribution and significance of each input parameter (material properties) on the output (group velocities). CNN with dual-branch feature fusion is implemented to solve two different inverse problems, i.e., (1) finding ply-layup sequence type and (2) identification of material properties. A classification-based network is implemented to classify polar representations into three important layup sequence types, whereas a regression-based network is used to map polar representations into six different material properties. Both networks have shown exceptional prediction and generalization capabilities on different datasets. The time taken to identify the layup sequence type and material properties are of the order of milliseconds. This additional advantage can be used for real-time implementation to understand the degradation of material properties as well as the rapid non-destructive measurement of material properties in mass-production systems. Various machine learning algorithms are applied to the featurized dataset and compared against our methodology. It is seen that our proposed approach performs much better and surpasses them.

\section{Data Availability}
The raw/processed data required to reproduce these findings cannot be shared at this time as the data also forms part of an ongoing study.

\section{Acknowledgement}
J.S. acknowledge funding from the Accelerated Materials Development for Manufacturing Program at A*STAR via the AME Programmatic Fund by the Agency for Science, Technology and Research under Grant No. A1898b0043.

\section{Disclosure statement}
No potential conflict of interest was reported by the authors.
%%%%%%%%%%%%%%%%%%%%%%%%%%%%%%%%%%%%%%%%%%%%%%%%%%%%
\bibliography{mybibfile_cs}
\end{document}